\documentclass[12pt,preprint]{emulateapj}
\usepackage{graphicx}
\usepackage{epsfig} 
\usepackage{multirow} 
\usepackage[usenames]{color}

\def\gsim{\;\rlap{\lower 2.5pt \hbox{$\sim$}}\raise 1.5pt\hbox{$>$}\;} 
\def\lsim{\;\rlap{\lower 2.5pt \hbox{$\sim$}}\raise 1.5pt\hbox{$<$}\;}

\newcommand{\atlas}{ATLAS$^{\rm 3D}\,$}
\newcommand{\kms}{{\rm \, km~s^{-1}}}
\newcommand{\ho}{\ensuremath{H_0}}
\newcommand{\mdot}{M_\odot}
\newcommand{\msun}{\ensuremath{M_{\odot}}}
\newcommand{\mbh}{\ensuremath{M_\mathrm{BH}}}
\newcommand{\mldyn}{\ensuremath{(M^*/L)_{\rm dyn}}}
\newcommand{\mlpop}{\ensuremath{(M^*/L)_{\rm pop}}}

\newcommand{\sers}{S{\'e}rsic}

\newcommand{\sig}{\ensuremath{\sigma}}

\shorttitle{The MASSIVE Survey}
\shortauthors{Ma et al.}

\begin{document}

\title{The MASSIVE Survey -- I. A Volume-Limited Integral-Field Spectroscopic Study
of the Most Massive Early-Type Galaxies within 108 Mpc}

\author{Chung-Pei Ma\altaffilmark{1}, Jenny E. Greene\altaffilmark{2},
  Nicholas McConnell\altaffilmark{3}, Ryan Janish\altaffilmark{4}, John
  P. Blakeslee\altaffilmark{5}, Jens Thomas\altaffilmark{6} and Jeremy
  D. Murphy\altaffilmark{2}}
\affil{\altaffilmark{1} Department of Astronomy, University of California, Berkeley, CA 94720, USA; cpma@berkeley.edu \\
  \altaffilmark{2} Department of Astrophysical Sciences,Princeton University, Princeton, NJ 08544, USA \\
  \altaffilmark{3} Institute for Astronomy, University of Hawaii at Manoa, Honolulu, HI 96822, USA \\
  \altaffilmark{4} Department of Physics, University of California, Berkeley, CA 94720, USA \\
  \altaffilmark{5} Dominion Astrophysical Observatory, NRC Herzberg
  Institute of
  Astrophysics, Victoria, BC V9E\,2E7, Canada\\
  \altaffilmark{6} Max Planck-Institute for
  Extraterrestrial Physics, Giessenbachstr. 1, D-85741 Garching, Germany\\
}

\label{firstpage} 
\begin{abstract}
  Massive early-type galaxies represent the modern-day remnants of the
  earliest major star formation episodes in the history of the universe.
  These galaxies are central to our understanding of the evolution of
  cosmic structure, stellar populations, and supermassive black holes, but
  the details of their complex formation histories remain uncertain.  To
  address this situation, we have initiated the MASSIVE Survey, a
  volume-limited, multi-wavelength, integral-field spectroscopic (IFS) and
  photometric survey of the structure and dynamics of the $\sim100$ most
  massive early-type galaxies within a distance of 108 Mpc.  This survey
  probes a stellar mass range $M^* \ga 10^{11.5} M_\odot$ and diverse
  galaxy environments that have not been systematically studied to date.
  Our wide-field IFS data cover about two effective radii of individual
  galaxies, and for a subset of them, we are acquiring additional IFS
  observations on sub-arcsecond scales with adaptive optics.  We are also
  acquiring deep $K$-band imaging to trace the extended halos of the
  galaxies and measure accurate total magnitudes.  Dynamical orbit modeling
  of the combined data will allow us to simultaneously determine the
  stellar, black hole, and dark matter halo masses.  The primary goals of
  the project are to constrain the black hole scaling relations at high
  masses, investigate systematically the stellar initial mass function and
  dark matter distribution in massive galaxies, and probe the late-time
  assembly of ellipticals through stellar population and kinematical
  gradients.  In this paper, we describe the MASSIVE sample selection,
  discuss the distinct demographics and structural and environmental
  properties of the selected galaxies, and provide an overview of our basic
  observational program, science goals and early survey results.
\end{abstract}
\keywords{galaxies: elliptical and lenticular, cD
--- galaxies: evolution
--- galaxies: kinematics and dynamics
--- galaxies: stellar content
--- galaxies: structure
--- dark matter}
\maketitle

\section{Introduction}\label{sec:intro}

The most massive early-type galaxies in the local universe are powerful
probes of galaxy evolution. They formed most of their stars rapidly at
redshifts $z > 2$ (e.g., \citealt{blakesleeetal2003, thomasetal2005}) but
have grown in number and size by a factor of two or more since $z \approx
1$ (e.g., \citealt{daddietal2005,trujilloetal2006,faberetal2007,
  vanderweletal2008,damjanovetal2009,cappellarietal2009,
  vandokkumetal2010,vandesandeetal2011}), probably in large part through
dissipationless merging and accretion (e.g., \citealt{deluciaetal2006,
  boylankolchinetal2006, naabetal2009, kormendybender2009, oseretal2010,
  thomasetal2014}).  They contain nuclear black holes whose masses \mbh\ are
correlated with properties of the stellar bulge (e.g.,
\citealt{magorrianetal1998, ferraresemerritt2000, gebhardtetal2000,
  tremaineetal2002, marconihunt2003, haringrix2004, gultekinetal2009,
  beifiorietal2012, mcconnellma2013, kormendyho2013}). These scaling
relations between black holes and their host galaxies imply co-evolution
between the two components over the lifetime of a galaxy, but the detailed
mechanisms remain uncertain.

Integral field spectroscopy (IFS) over a wide radial range provides an
effective tool to study the spatial and kinematic structure, star formation
histories, and stellar and dark matter halo masses of local galaxies.
While many IFS surveys are ongoing, such as VENGA/VIXENS
\citep[][]{blancetal2013}, CALIFA \citep{sanchezetal2012}, SLUGGS
\citep[][]{brodieetal2014}, and eventually MaNGA (Bundy et al. in
preparation) and SAMI \citep{croometal2012}, none of them probes the
volume, mass range, or spatial scales required to systematically study the
most massive elliptical galaxies, a regime that is critical for
understanding the assembly of galaxies and supermassive black holes.  The
\atlas project \citep{cappellarietal2011} surveyed 260 galaxies within 42
Mpc.  Because of their relatively small survey volume, only a handful of
galaxies had stellar masses $M^* \gsim 10^{11.5}\mdot$.  Their field-of-view
of $33\arcsec\times 41\arcsec$ provided coverage within one half-light
radius of most of their sample galaxies.

In this paper we describe MASSIVE, a volume-limited, multi-wavelength,
spectroscopic and photometric survey of the most massive galaxies in the
local universe.  The sample includes 116 candidate galaxies in the northern
sky with distance $D < 108$ Mpc and absolute $K$-band magnitude
$M_K{\,<\,}{-}25.3$, corresponding to stellar masses $M^* \gsim
10^{11.5}M_\odot$.  MASSIVE is designed to address a wide range of
outstanding problems in elliptical galaxy formation, including the
variation in dark matter fraction and stellar initial mass function (IMF)
within and among early-type galaxies, the connection between black hole
accretion and galaxy growth, and the late-time assembly of galaxy
outskirts.  We combine comprehensive ground-based NIR imaging with IFS data
to measure stellar populations and kinematics out to $\sim 2$ effective
radii.  Using the Mitchell Spectrograph \citep[formerly called
VIRUS-P;][]{hilletal2008a} at McDonald Observatory, we cover a
$107\arcsec\times 107\arcsec$ field of view with 4\arcsec\ fibers.  Thus,
we are sensitive to low surface-brightness emission in the outer parts of
the galaxies \citep[e.g.,][]{murphyetal2011,adamsetal2012, greeneetal2012}.

For a subset of galaxies, we are obtaining additional adaptive-optics
assisted IFS data to map the stellar kinematics on $\sim 100$ pc scales,
within the sphere of influence of nuclear black holes.  The high-resolution
data are required to detect the gravitational effect of the supermassive black
holes on the stellar orbits.  These data alone, however, are insufficient for
removing the degeneracy among the dark matter halo, the stellar
mass-to-light ratio, and the central black hole mass
\citep{gebhardtthomas2009}.  We therefore combine the high-resolution and
wide-field IFS data for simultaneous modelling of the three mass components
\citep[e.g.,][]{schulzegebhardt2011, mcconnelletal2011a,
  mcconnelletal2011b, mcconnelletal2012, ruslietal2013, thomasetal2014}.

We are also acquiring deep $K$-band data at UKIRT and CFHT to measure the
extended stellar halos of these luminous galaxies and determine more
accurately their $K$-band magnitudes.

The wide-field IFU portion of the survey is currently complete at $M_K <
-25.7$ and 75\% complete at $M_K < -25.5$.  Seven galaxies in the survey
have published values of black hole mass (Sec~3.6); 15 additional galaxies
have existing or incoming high-resolution kinematic data.  Deep $K$-band
photometry has been obtained for 45 galaxies thus far.

The selection of the galaxy sample for the MASSIVE survey is described in
Section~2.  Since this survey is volume-limited and defined by the stellar
masses of the galaxies via their $K$-band luminosities, we discuss in
detail the determinations of distance and absolute $K-$band magnitude.
Basic properties of the survey galaxies such as stellar mass, size,
velocity dispersion, shape, color, and central black holes are presented in
Section~3.  We illustrate the distinct demographics of these galaxies and
compare their locations in parameter space with lower-mass early type
galaxies.  In Section~4, we investigate the larger-scale environments of
these massive galaxies using three 2MASS-selected galaxy group catalogs
within the local volume.  Our observing strategies with large-format IFS,
AO-assisted IFS, and deep $K$-band imaging are discussed in Section~5.
Sample spectra from the Mitchell IFS and kinematic maps of NGC 1600 are
shown.  We discuss the primary science goals of the survey 
and present early science results in Section~6.

Appendix~A tabulates the 116 candidate galaxies in the MASSIVE survey and their
key physical properties.  Appendix~B provides a montage of the 78 MASSIVE
galaxies with SDSS photometry.

We assume $H_0=70 \kms$ Mpc$^{-1}$ throughout the paper.

\section{Sample Selection}\label{sec: sample}

\subsection{Overview}

The main selection criteria of our survey are summarized in Table~1.  The
survey volume of radius $D<108$ Mpc is chosen to be large enough to
encompass the Coma cluster.  This volume is more than an order of magnitude
larger than that probed by \atlas, enabling us to obtain a statistical
sample of early-type galaxies at the highest end of the galaxy mass
function.  The corresponding redshift limit is $cz < 7560 \kms$ or $z <
0.025$ (for $H_0=70 \kms$ Mpc$^{-1}$).

Within the survey volume, our goal is to select galaxies above a given
total stellar mass.  Since selection based on $K$-band luminosities is
close to a stellar mass selection, particularly for these red galaxies, we
use the near-infrared $K$-band magnitude from the Extended Source Catalog (XSC;
\citealt{jarrettetal2000}) of the Two Micron All Sky Survey (2MASS;
\citealt{skrutskieetal2006}).  This catalog contains photometric
measurements in the $J$, $H$, and $K$ bands of $\sim 1.6\times 10^6$
objects with $K \le 13.5$ mag.  The data have a mean photometric accuracy
better than 0.1 mag and are mostly unaffected by interstellar extinction
and stellar confusion, although the 2MASS luminosities may be
systematically underestimated for very extended objects (see Sec~\ref{sec:K}).

Peculiar velocities add uncertainties to the determination of distances and
absolute magnitudes, and consequently the selection of our sample.  We use
the 2MASS Galaxy Redshift Survey (2MRS; \citealt{huchraetal2012}) and the
group catalog based on 2MRS \citep{crooketal2007} to correct the radial
velocity-derived distances.  We begin with an initial velocity cutoff
corresponding to a redshift-distance of 140 Mpc and $M_K < -22$ and correct
for peculiar velocities for galaxies in the 2MRS group catalog (see
Sec~\ref{sec:distance}).
We then select those galaxies with $D < 108$ Mpc, $M_K < -25.3$ mag,
declination $\delta > -6^\circ$, and galactic extinction $A_V < 0.6$.

Finally, we restrict our sample to galaxies classified as elliptical or S0
in the HyperLeda database\footnote{http://leda.univ-lyon1.fr}
\citep{patureletal2003}.  We remove 14 galaxies from the sample because
their photometry is compromised by either a foreground star or a companion
galaxy, and the stellar mass is likely to be overestimated (see Sec~2.5).  We do not
remove any galaxies based on their size on the sky; in practice, most
galaxies in the survey have effective radii larger than 10\arcsec\ (listed
in Table~\ref{big_table}; see discussion in Sec~\ref{sec:size}) and are
therefore well-resolved by the 4\arcsec\ fibers of the Mitchell
Spectrograph.

\begin{table}
\begin{center}
  \caption{Selection criteria for MASSIVE galaxies}
  \begin{tabular}{ll}
   \hline
   Distance & $D < 108$ Mpc \\
   Absolute $K$ magnitude & $M_K < -25.3$ \\
   Declination & $\delta > -6^\circ$ \\
   Galactic extinction & $A_V < 0.6$ \\
   Morphology & E and S0 \\
   \hline
  \end{tabular}
\end{center}
\vspace{0.3in}
\end{table}

This set of selection criteria results in 116 candidate galaxies listed in
Table~\ref{big_table}.  Among these, 72 galaxies have $M_K < -25.5$ and $D
< 105$ Mpc and are our high priority targets.  We are obtaining deeper
$K$-band imaging to improve on the relatively shallow photometry provided
by 2MASS.  The more robust measurements of the total $K$-band magnitude for
our candidate galaxies will help alleviate the uncertainties near the
magnitude and distance cutoffs and sharpen the survey boundaries and the
final sample size.

Below we describe the key selection criteria in more detail.

\subsection{Distance}
\label{sec:distance}

We need accurate distance estimates to determine the absolute $K$-band
magnitudes, the volume cutoff, and the measurements of \mbh\ and $M^*$.
Only 9 galaxies in our survey have existing distances measured from the
surface-brightness fluctuation (SBF) method
\citep[e.g.,][]{blakesleeetal2009, blakesleeetal2010} for either the
individual galaxies or the groups in which they reside.  Among these, three
are in the Virgo cluster (NGC 4472, 4486, 4649) at 16.7 Mpc
\citep{blakesleeetal2009}, four are in the Coma cluster (NGC 4816, 4839,
4874, 4889) at 102.0 Mpc \citep{blakeslee2013}, and two are in the Perseus
group (NGC 7619 and 7626) at 54.0 Mpc \citep{cantielloetal2007}.  We adopt
SBF distances for these 9 galaxies.

For the rest of the sample, we assign the distance in one of two ways,
depending on whether or not a galaxy is identified as belonging to a group.
For galaxies in groups, we correct for local peculiar velocities using
group-corrected redshift distances; for galaxies not in groups, we use
redshift distances based on radial velocities corrected with a flow model,
as described below.

To determine group membership, we use the catalog of galaxy groups
constructed from the friends-of-friends (FOF) algorithm applied to 2MRS
\citep{crooketal2007}.  The 2MRS contains follow-up spectroscopic data and
redshifts for a subset of 43,533 galaxies in 2MASS.  It is 97.6\% complete
down to $K=11.75$ mag over 91\% of the sky.  The median uncertainty in the
radial velocities for galaxies with absorption-line spectra is 29 and $41
\kms$, respectively, for the two main spectrographs used in the survey.
\citet{crooketal2007} present a catalog of galaxy groups using the 2MRS
redshifts.  It is complete to a limiting radial velocity of $10^4 \kms$.  The
high-density-contrast (HDC) catalog in this work provides galaxy membership
in groups that have a density contrast of 80 or more, corresponding to FOF
linking parameters of 0.93 Mpc (for $h=0.7$) in the transverse directions
and $350 \kms$ along the line of sight.

For galaxies that reside in HDC groups with three or more members, we use
the mean group distance from the HDC catalog, converted from $H_0=73 \kms$
Mpc$^{-1}$ to our adopted value of $70 \kms$ Mpc$^{-1}$.  The group
distance is determined using velocities from the flow model of
\citet{mouldetal2000} to account for the most obvious local distortions and
large-scale flows.  The model first converts from the heliocentric frame to
the Local Group frame, and then adjusts the redshift-inferred distances of
galaxies near the Virgo Cluster, Shapley Supercluster, and the Great
Attractor region.\footnote{However, we use the SBF distance for the Virgo
  galaxies, and none of our galaxies are in Shapley or the Great Attractor
  regions, as these are in the southern sky.}  Then, the Local Group frame
velocities for all galaxies are corrected for the estimated gravitational
pull of the Virgo, the Great Attractor, and Shapley mass concentrations.

For galaxies not residing in any HDC group, we assign the distances using
velocities from the same flow model (as provided by NED and converted to
our \ho) to ensure that the distances for group and field galaxies in our
survey are computed in the same rest frame.

While we have used the best available distance measurements (listed in
Table~\ref{big_table}), uncertainties will unavoidably cause a small
fraction of the galaxies near our mass and distance cutoffs to move into
and out of the sample.  We have attempted to quantify the outstanding
distance uncertainties by comparing our adopted distances with the
redshift-independent distances tabulated by NED for 39 objects in our
sample.  We find the mean offset to be 1.5 Mpc, but $\sim 20$\% of the
cases differ by $> 10$~Mpc. Since galaxy populations over a few tens of Mpc
or a few tenths of a magnitude are not expected to change, we do not
anticipate the uncertainty in the exact membership near our survey
boundaries to impact our results.  Our dynamical measurements of \mbh\ and
$M^*$, however, do depend on the distance linearly, and all current such
measurements are affected by this uncertainty.  Significant improvements
can be achieved with more SBF data with the {\it Hubble Space Telescope}.

\subsection{$K-$band magnitude}

\label{sec:K}

The 2MASS XSC database provides a variety of magnitude measurements for
each extended source in the $J$, $H$, and $K$ bands.  To determine each
galaxy's absolute $K$-band luminosity, we begin with the ``total''
extrapolated $K-$band magnitude (XSC parameter k\_m\_ext), which is
measured in an aperture consisting of the isophotal aperture plus the
extrapolation of the surface brightness profile based on a single \sers\
fit to the inner profile \citep{jarrettetal2003}.
We compute the absolute $K$-band magnitudes using
\begin{equation}
M_K=K - 5 \log_{10} D - 25-0.11 A_V \,,
\label{MK}
\end{equation}
where $K$ is given by k\_m\_ext, and $D$ is the distance in Mpc described
in Sec~\ref{sec:distance}.  We use galactic extinction $A_V$ (Landolt $V$)
from \citet{schlaflyetal2011} and the reddening relation of
\citet{fitzpatrick1999} with $R_V=A_V/E_{B-V}=3.1$.  The values of $K$,
$A_V$, and $M_K$ for all galaxies are listed in Table~\ref{big_table}.
These $K-$band magnitudes form the basis for our selection and are used to
estimate stellar masses (\S~\ref{sec:mstar}).

The relatively shallow photometry provided by 2MASS XSC (the 1-$\sigma$
sky noise in $K$ is 20.0 mag arcsec$^{-2}$) has led to some concerns that the
$K$-band luminosities of massive galaxies are underestimated by 2MASS
(e.g., \citealt{laueretal2007, schombertsmith2012, kormendyho2013}).  When
the radial range is too small to provide an accurate Sersic index for the
light profiles of early-type galaxies, their total luminosities can be
particularly biased low.

To assess the impact of potential biases in 2MASS $K$-band magnitudes on
our galaxy selection, we examine the sample of 219 early-type galaxies
targeted for an HST imaging study in \citet{laueretal2007}.  We use
$V-K=2.98$ \citep{kormendyho2013} to transform the $V$-band luminosities
(largely based on the RC3 Catalog) in their sample to the $K$-band.  A
total of 31 galaxies in this sample lie within our survey volume of $D <
108$ Mpc and above the luminosity cut of $M_K = -25.3$ mag; among these, 18
have $\delta > -6^\circ$ and would belong to our survey if the deeper
$V$-band photometry and $V-K$ relation were used to select bright
galaxies.\footnote{We exclude IC 1565, which is at a distance of $\sim150$
  Mpc according to NED; it is incorrectly listed as 38.2 Mpc in
  \citet{laueretal2007}.}  We find all 18 to be in our MASSIVE sample; our
$K$-band selection therefore does not appear to be much affected by
potential systematic underestimates in the 2MASS $K$-band magnitude
according to this test.


We are acquiring deep wide-field $K$-band photometry (Sec~\ref{sec:deepK})
for more robust measurements of the total magnitude for the galaxies in our
sample.  M87, for instance, has $M_K=-25.31$ mag according to the 2MASS XSC
and is only slightly above our magnitude cut.  Analysis of deeper
photometric data, however, finds $M_K = -26.08$ mag, a factor of two more
luminous \citep{lasker2014a}.  Although M87 is likely to be a worst case
because of its angular extent on the sky, this large discrepancy
underscores the need for deeper $K$-band imaging data.  Our dataset will
also reduce the uncertainties near our magnitude and distance cutoffs and
help refine the final sample selection for IFS observations.

\subsection{Parameter Space}

\begin{figure}
\vspace{-.8in}
\hspace{-0.2in}
\includegraphics[width=4.1in]{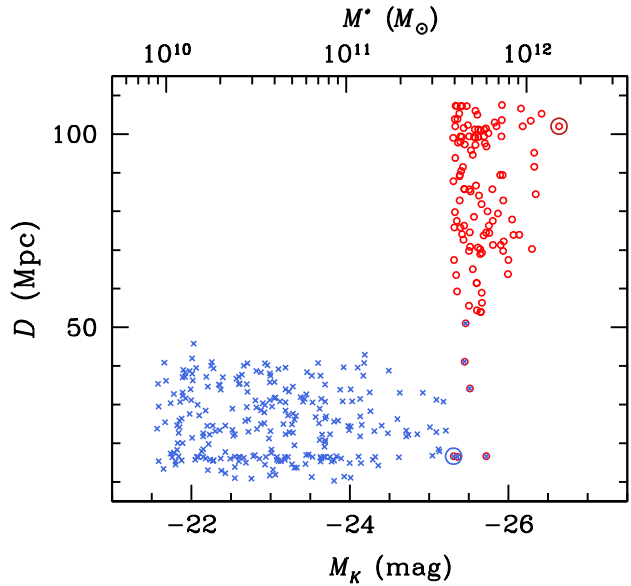}
\vspace{-1.2in}
\caption{Distance and absolute $K$-band magnitude of galaxies in the
  MASSIVE survey (red circles) and \atlas survey (blue crosses). Stellar
  masses estimated from Equation~(\ref{MstarMK}) are also shown.  Only 6 galaxies in
  \atlas are luminous enough to pass our $K$-band magnitude cut.  The rest
  of our sample all lie beyond the volume limit of 42 Mpc surveyed by
  \atlas.  The big circles indicate NGC 4889 (red) and M87 (blue).  }
\vspace{0.3in}
\label{mass_distance}
\end{figure}

Figure~\ref{mass_distance} highlights the distinct parameter space in
distance and stellar mass occupied by MASSIVE galaxies.  Only 6 galaxies in
this survey were included in \atlas; three are in the Virgo Cluster: NGC
4486 (M87), NGC 4472 (M49), NGC 4649 (M60); the others are NGC 5322, NGC
5353, and NGC 5557.  The larger survey volume (by more than a factor of 15)
allows us to sample the galaxy mass function at $M^* \ga 10^{11.5} \mdot$.

The ongoing CALIFA survey will target $\sim 600$ diameter-selected local
galaxies with major axes between 45\arcsec\ and 80\arcsec\ (in the SDSS
$r$-band) and a redshift range of $0.005 < z < 0.03$.  About 1/3 of the
galaxies are expected to be bulge-dominated
\citep{sanchezetal2012}. Despite the selection on galaxy sizes, the CALIFA
sample is shown to be representative of galaxies with $ 10^{9.7} < M^* <
10^{11.44} \mdot$ \citep{walcheretal2014}, complementary to our mass
selection of $M^* \ga 10^{11.5} \mdot$.


\begin{figure}
\vspace{-0.8in}
\hspace{-.2in}
\includegraphics[width=4.1in]{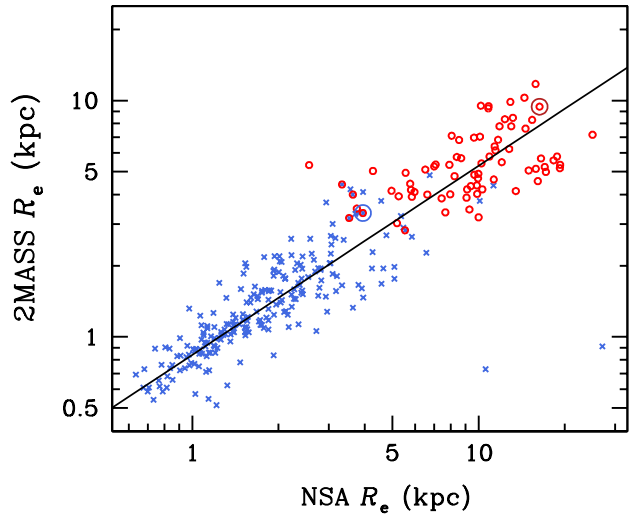}
\vspace{-1.2in}
\caption{Comparison of infrared (from 2MASS) and optical (from NSA) galaxy
  sizes for the MASSIVE (red circles) and \atlas (blue crosses) surveys.
  77 galaxies in the MASSIVE sample are listed in the NSA. The 2MASS $R_e$
  is systematically lower than the NSA values.  The best-fit relation
  (black line) is given in Equation~(\ref{re-re}).  The big circles indicate NGC
  4889 (red) and M87 (blue).  }
\label{size_size}
\vspace{0.22in}
\end{figure}

\subsection{Skipped Targets}
\label{sec:skipped}

A total of 14 galaxies pass our selection criteria but are in the field of
view of a bright star or have a companion or interacting galaxy.  We list
these galaxies here for completeness, but we do not include them in our
candidate list because their photometry is likely to be contaminated by the
near neighbor and the luminosities may be overestimated.  Among the 14
galaxies, 4 have nearby stars: NGC 2974,\footnote{We note that NGC~2974 was
  included in \atlas but was assigned $M_K=-23.62$ mag, much fainter than
  our survey magnitude cut and the 2MASS XSC value $K=6.24$ mag and a
  corresponding $M_K=-26.16$ mag.  The 2MASS isophotal radius is 71" for
  this galaxy, enclosing the $V=9.2$ mag bright star BD-03 2751 that is
  43'' away.  The 2MASS magnitude for NGC 2974 is therefore highly
  contaminated.} NGC 6619, IC 947, UGC 11950; the other 10 have interacting
or close companion galaxies: NGC 71, NGC 750, NGC 1128, NGC 4841A, NGC
5222, NGC 7318, PGC 27509, PGC 93135, UGC 2759, UGC 12591.

One exception is the close galaxy pair NGC~545/547.  NGC~547 is in 2MASS
and passes our selection cut.  NGC~545 is not in 2MASS, but it is listed as
the BCG of Abell~194 with $M_V=-22.98$ mag \citep{laueretal2007}.  We
include both galaxies in our sample.

We exclude NGC~1275, the central galaxy in the Perseus Cluster, since it is
a complex post-merger system \citep[e.g.,][and references
therein]{canningetal2014}.

\section{Galaxy Properties}

\subsection{Stellar Mass}
\label{sec:mstar}

A major goal of this survey is to obtain measurements of the stellar
mass-to-light ratio of massive early-type galaxies from both dynamical
modeling and stellar population synthesis modeling of the IFS kinematic
data.  In the interim, we estimate the stellar mass of the survey galaxies
using a conversion between $K$-band luminosity and stellar mass for
early-type galaxies in the \atlas sample \citep{cappellari2013}:
\begin{equation}
  \log_{10} (M^*) = 10.58 - 0.44 (M_K + 23)\,. 
\label{MstarMK}
\end{equation}
The relation is fitted between total extinction-corrected 2MASS $K$-band
magnitudes and dynamical stellar masses from Jeans Anisotropic MGE (JAM),
where MGE is the Multi-Gaussian Expansion method \citep{emsellemetal1994}.
This scaling naturally incorporates any potential IMF changes as a function
of mass. Upon the completion of the MASSIVE survey, we will be able to test
the validity of this conversion for the mass range $M^* \ga 10^{11.5}
\mdot$, which is currently uncalibrated at this high mass.

\subsection{Galaxy Size}
\label{sec:size}

\begin{figure}
\vspace{-1.2in}
\hspace{-.2in}
\includegraphics[width=4.1in]{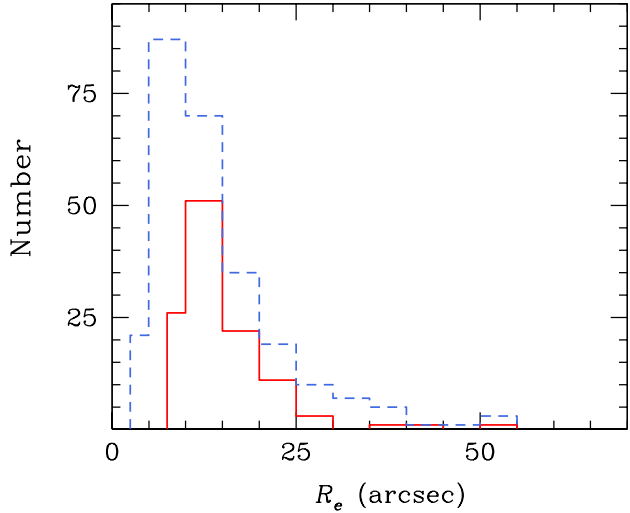}
\vspace{-1.2in}
\caption{Distribution of 2MASS angular sizes for the MASSIVE (red solid)
  and \atlas (blue dashed) galaxies. 
  The Mitchell IFS has a FOV of $107\arcsec\times 107\arcsec$ and probes up
  to $\sim 2 R_e$ for most MASSIVE galaxies.  The FOV of Sauron/\atlas is
  $33\arcsec\times 41\arcsec$.}
\label{ang_size}
\vspace{0.3in}
\end{figure}

The 2MASS XSC catalog lists a variety of measurements for galaxy sizes.
For ease of comparison, we adopt a similar definition of the effective
radius $R_e$ as \citet{cappellarietal2011}.  Their $R_e$ is based on the
half-light radius from XSC (parameters j\_r\_eff, h\_r\_eff, and
k\_r\_eff).  This radius is derived from the 2MASS surface brightness
profile in each band as the value of the semi-major axis of the ellipse
that encloses half of the total light.  We assign each galaxy a 2MASS $R_e$
using the median value in the three bands:
\begin{equation}
  R_e = {\rm median}({\rm j\_r\_eff, h\_r\_eff, k\_r\_eff}) \sqrt{{\rm sup\_ba}}\,,
\label{eq:reff}
\end{equation}
where the parameter sup\_ba is the minor-to-major axis ratio measured from the
2MASS 3-band co-added image at the $3\sigma$ isophote.\footnote{We use
  sup\_ba instead of the $K$-band axis ratio k\_ba adopted by
  \citet{cappellarietal2011} because sup\_ba is measured from the higher S/N
  combined images and is listed to 3 rather than 1 decimal precision in
  2MASS XSC.}  This factor is included here to convert the semi-major axis
into the radius of the circle with the same area.
\citet{cappellarietal2011} found the 2MASS $R_e$ for \atlas galaxies to
correlate well with the optical $R_e$ from the RC3 catalog \citep{RC31991}
with an rms scatter of 0.12 dex, but the 2MASS radii were smaller by an
overall factor of $\approx 1.7$, presumably because 2MASS is shallow
\citep[see also][]{laueretal2007}.

Here we compare the 2MASS $R_e$ with the optical sizes from the NASA-Sloan
Atlas (NSA), which in turn is based on the SDSS DR8 spectroscopic catalog
\citep{yorketal2000,aiharaetal2011}.  This version of the SDSS photometric
catalog has a revised sky subtraction designed specifically to mitigate
known galaxy size measurement problems for large galaxies
\citep[][]{desrochesetal2007, blantonetal2011}. 
The NSA provides a unified analysis of local galaxies within $\sim 200$ Mpc. 
A total of 77 MASSIVE galaxies are in the NSA.  For the optical
$R_e$, we use the 50\% light radius from a 2-dimensional \sers\ fit along
the major axis (NSA parameter SERSIC\_TH50).  The \sers\ indices from the
NSA fits range from $n=2$ to the maximum allowed $n=6$.

The values of 2MASS and available NSA radii are listed in
Table~\ref{big_table}.  Figure~\ref{size_size} compares the physical $R_e$
from 2MASS and NSA for galaxies in the MASSIVE and \atlas surveys. 
The best-fit relation (black line) is
\begin{equation}
 \log_{10} R_e^{\rm 2MASS} = 0.80 \log_{10} R_e^{\rm NSA}- 0.076 \,,
\label{re-re}
\end{equation}
where the radii are in kpc.  At $\sim 1$ kpc, the NSA $R_e$ is a factor of
$\sim 1.2$ larger than the 2MASS $R_e$.  At $\sim 10$ kpc, the offset
increases to a factor of $\sim 1.8$.  This difference underscores the need
for deeper photometry, particularly in the $K$-band and for massive
galaxies.

Figure~\ref{ang_size} shows the distributions of the 2MASS angular $R_e$
for galaxies in the two surveys.  Most MASSIVE galaxies are in the range of
$\sim 10\arcsec$ to 50\arcsec\ , comparable to those of \atlas galaxies.
Due to the larger distances, however, the physical sizes of MASSIVE
galaxies are $\sim 2$ to 5 times larger. This is consistent with the large
stellar masses of these galaxies.  The $107\arcsec \times 107\arcsec$ FOV
of our IFS covers up to $\sim 2R_e$ of most galaxies in the MASSIVE survey,
in comparison to the $33\arcsec \times 41\arcsec$ FOV of \atlas.

\begin{figure}
\vspace{-1.2in}
\hspace{-.2in}
\includegraphics[width=4.1in]{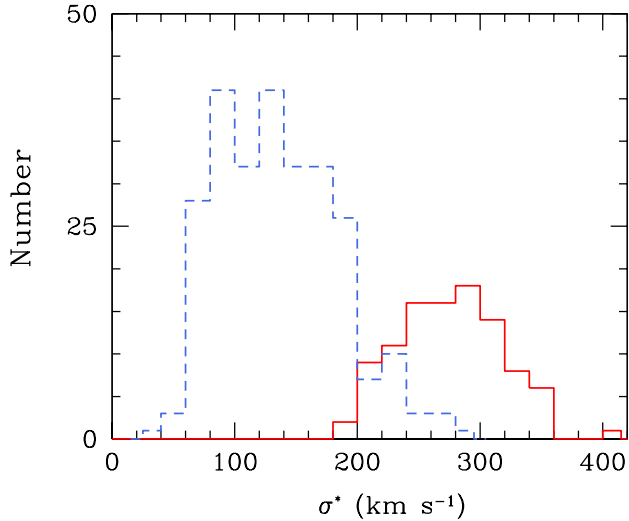}
\vspace{-1.2in}
\caption{Distribution of stellar velocity dispersion for the
  MASSIVE (red solid) and \atlas (blue dashed) galaxies.}
  \label{sigma_hist}
\end{figure}

\subsection{Stellar Velocity Dispersions}

\begin{figure*}
\vspace{-.8in}
\includegraphics[width=4.1in]{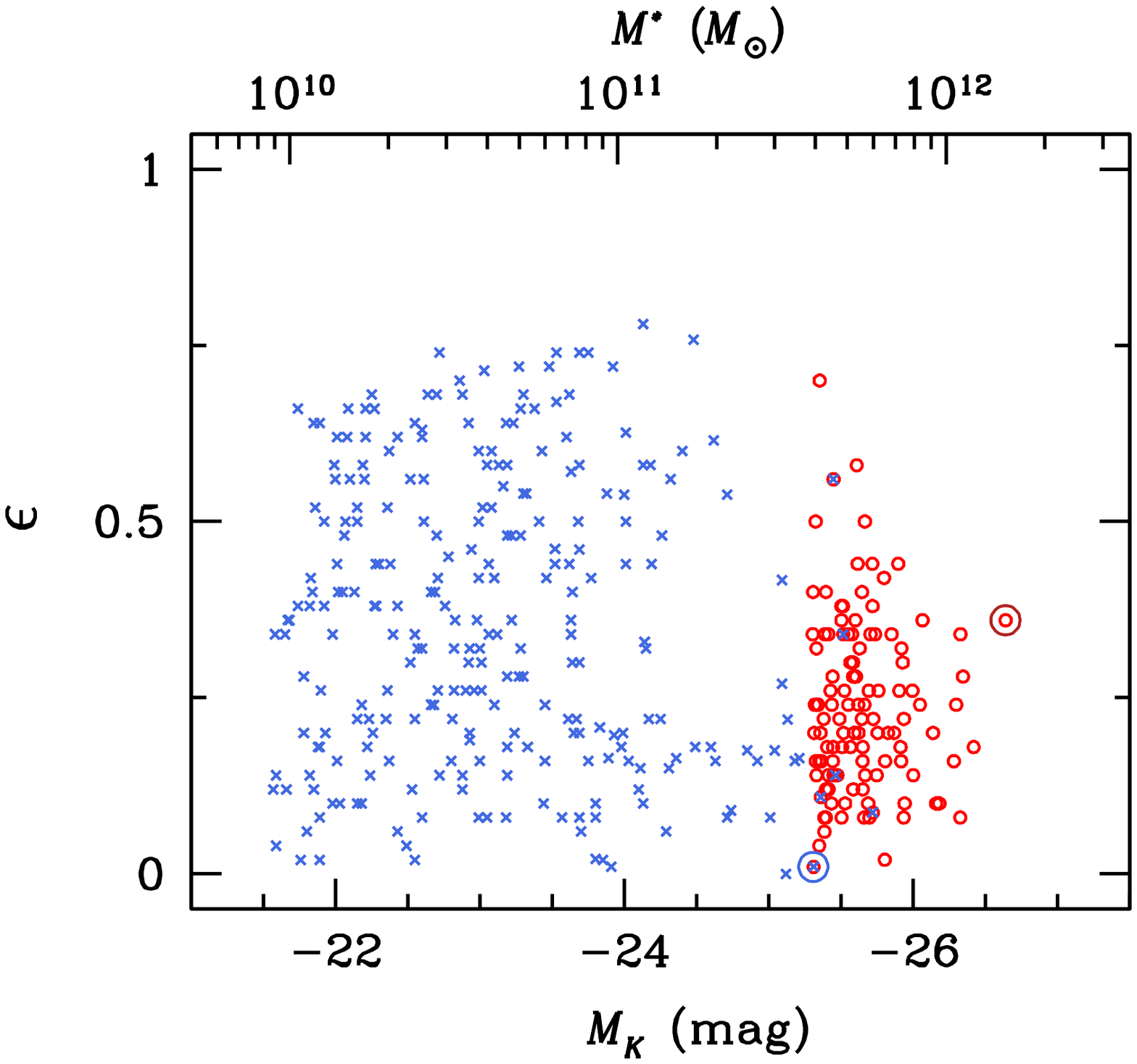}
\vspace{-1.2in}
\hspace{-.5in}
\includegraphics[width=4.1in]{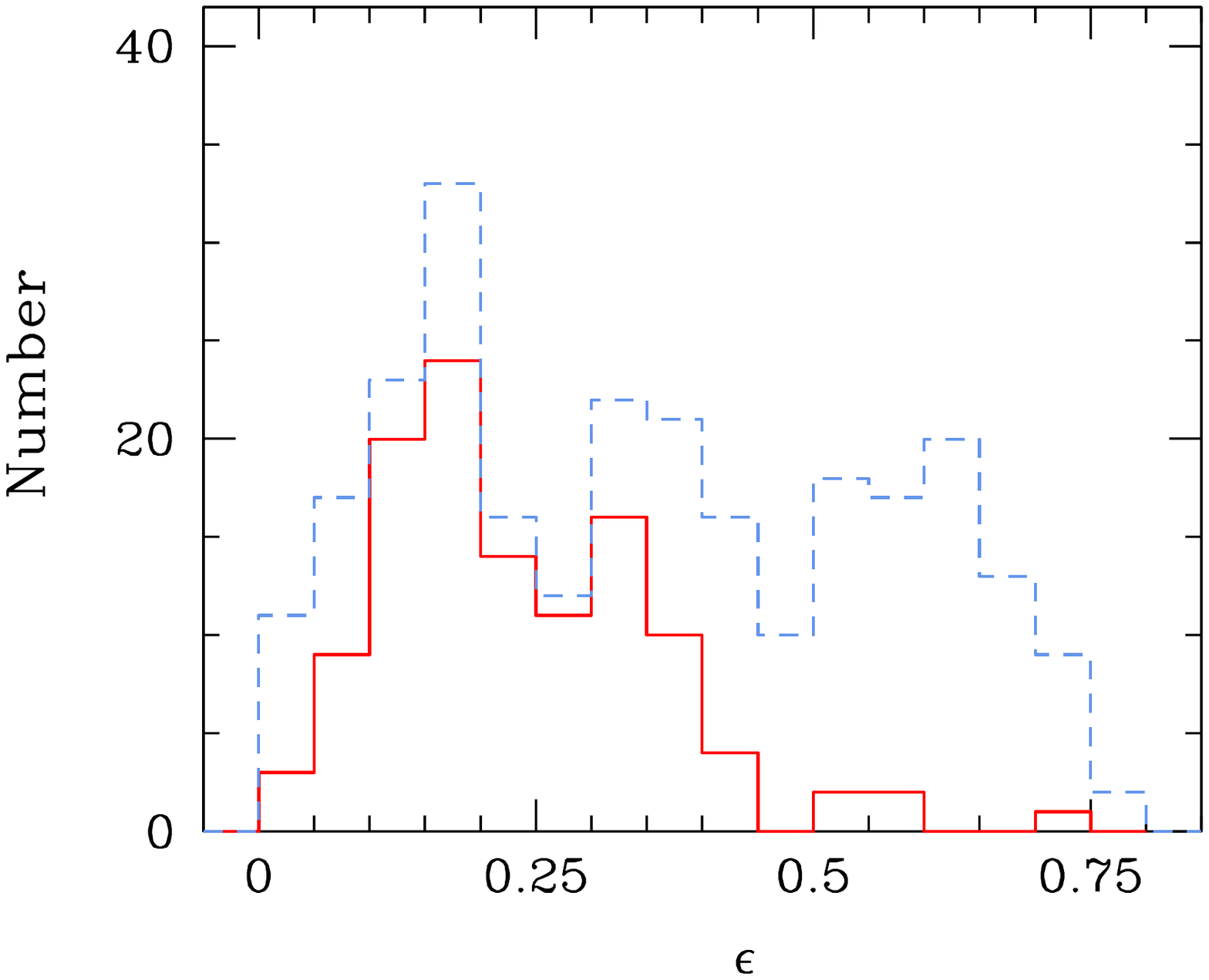}
\caption{Ellipticity versus $K$-band luminosity (left panel) and ellipticity
  distribution (right panel) for galaxies in the MASSIVE survey (red
  circles) and \atlas survey (blue crosses).  The ellipticity is $\epsilon
  = 1 - {\rm sup\_ba}$, where sup\_ba is the 2MASS XSC parameter for the
  minor-to-major axis ratio fit to their ``3-$\sigma$ super-coadd
  isophote.'' There is a dearth of high-$\epsilon$ galaxies in the MASSIVE
  sample.  The big circles in the left panel indicate NGC 4889 (red) and
  M87 (blue).  }
\vspace{0.3in}
\label{epsilon_mass}
\end{figure*}

A total of 98 MASSIVE galaxies have stellar velocity dispersion
measurements in the HyperLeda database \citep{patureletal2003}.  Among
those not in HyperLeda, five have $\sigma$ measurements in the SDSS
\citep{boltonetal2012}.  These 103 values are plotted in
Figure~\ref{sigma_hist} and listed in Table~\ref{big_table}; all other
available values of SDSS $\sigma$ are also listed.  The literature $\sigma$
values are measured over a range of radial apertures, so HyperLeda has
homogenized the measurements in a way designed to correct (on average) for
aperture effects as well as other differences in technique among different
studies (\citealt{prugnielsimien1996}; see also \citealt{ho2007}).  The
HyperLeda measurements are compiled as follows: (i) choose a subsample of
galaxies with three or more $\sigma$ measurements in the literature; (ii)
pick those that agree within $30 \kms$; and (iii) for each source, derive
offsets to match a gold sample of $\sigma$.  The final reported dispersion
is a weighted mean of scaled values.  The corrected velocity dispersions
from HyperLeda correspond to an aperture of 0.6 kpc.

The range of $\sigma$ (Figure~\ref{sigma_hist}) for our survey galaxies is
large, starting at $\sim 200 \kms$ up to $400 \kms$ for NGC 4889.  Two
galaxies have anomalous $\sigma$ in HyperLeda: NGC 4059 with $121 \kms$ and
NGC 4055 with $500 \kms$.  We replace them with $206 \kms$ and $270 \kms$,
respectively, from the NSA. Ultimately, our survey will produce
spatially-resolved 2-dimensional maps of velocities and will update the
$\sigma$ measurements.

\subsection{Shape}

The shapes, kinematics, and masses of early-type galaxies are closely
correlated.  Lower-mass elliptical galaxies tend to be fast rotators and
have higher ellipticities, whereas giant ellipticals rotate slowly and are
round and mildly triaxial (e.g., \citealt{binney1978,
  daviesetal1983,kormendybender1996,tremblaymerritt1996}).  It is therefore
interesting to examine the distributions in galaxy shapes for the MASSIVE
and \atlas\ samples.

Figure~\ref{epsilon_mass} compares the ellipticities, $\epsilon = 1 - {\rm
  sup\_ba}$, for galaxies in the two surveys, where sup\_ba is the 2MASS
XSC parameter for the minor-to-major axis ratio fit to their
``super-coadd'' isophote.  Only five MASSIVE galaxies have high
ellipticities with $\epsilon \ga 0.5$, in contrast to about a quarter of
the \atlas\ sample.  These five galaxies are all in the fainter half ($M_K
\gsim -25.7$ mag) of our sample.  
Our survey data will provide direct measurements of the spatial profile of
the rotation and $\epsilon$ of each galaxy and will allow us to quantify
the distributions of galaxy rotations and shapes at the highest masses.

\subsection{Color}

Galaxies in the MASSIVE survey are selected based on properties such as
luminosity and morphology but not color.  We quantify their color
distribution using the photometry for the 77 MASSIVE galaxies that are in
the NSA.  We find that the $u-r$ distribution is well-described as a
Gaussian with a mean color of $u-r = 2.7 \pm 0.06$ mag.  This level of
scatter is similar to those quoted in earlier work (e.g.,
\citealt{boweretal1992, bernardietal2003, blantonetal2005,
  eisenhardtetal2007}).  The small scatter in the color-magnitude relation
is most likely tied to a uniformly old age and a narrow range in stellar
metallicity for these galaxies.  The wide range of environments of our
galaxies (see Sec~\ref{sec:env}) will enable us to identify any potential
color differences among the most massive galaxies as a function of local
environments.

\subsection{Supermassive Black Holes}
\label{sec:SMBH}

Seven galaxies in our sample have published black hole masses in the
literature: the three Virgo galaxies NGC 4486 \citep{gebhardtetal2011,
  walshetal2013}, NGC 4472 \citep{ruslietal2013}, and NGC 4649
\citep{shengebhardt2010}; NGC 3842 and NGC 4889 \citep{mcconnelletal2011a,
  mcconnelletal2012}; NGC 7052 \citep{marelbosch1998}, and NGC 7619
\citep{ruslietal2013}.  These galaxies are located at the high end of
the \mbh\ -$M^*$ relation \citep{mcconnellma2013}, but due to the large
scatter in $\sigma$ vs $M^*$ (Figure~\ref{sigma_mass}), the high end of
the \mbh\ -$\sigma$ relation is populated by a mixture of these massive
galaxies and several others not massive enough to be in our survey.  Recent
efforts at measuring large \mbh\ have all targeted high-$\sigma$ galaxies
(e.g., \citealt{mcconnelletal2011a, mcconnelletal2012, vdboschetal2012, ruslietal2013}). 
Our survey will provide a complementary sample of \mbh\ in galaxies
selected based on high stellar mass.

\section{Galaxy Environments}
\label{sec:env}

In this section we investigate the larger-scale environments of galaxies in
the MASSIVE survey.  Massive early-type galaxies are commonly assumed to be
located at or near the centers of galaxy groups or clusters.  Our survey
targets the most massive galaxies within a $\sim 100$ Mpc volume.  Where do
these $M^* \ga 10^{11.5}\mdot$ galaxies reside?  Below we quantify their
environments using three group catalogs constructed from galaxy redshift
surveys of the local volume.

\subsection{2MRS Group Catalog}

\citet{crooketal2007, crooketal2008} presents a redshift-limited catalog of
groups for the galaxies with $K < 11.25$ mag in 2MASS XSC.  The FOF
algorithm with two sets of linking parameters is used to create two group
catalogs of differing density contrasts.  The high-density-contrast (HDC)
catalog lists galaxy membership in groups that have a density contrast of
80 or more, corresponding to linking parameters of $350 \kms$ along the
line of sight and 0.89 Mpc in the transverse directions.  The
low-density-contrast (LDC) catalog is constructed with larger linking
lengths of $399 \kms$ and 1.63 Mpc, corresponding to a density contrast of
12 or more.

The exact membership of groups in any group/cluster catalog depends on the
algorithm and linking parameters used to construct the catalog.  All
galaxies assigned to groups in the HDC catalog are also assigned to groups
in the LDC catalog, but the converse is not true.  The larger linking
lengths used in LDC tend to merge smaller groups and generate more extended
structures, whereas large structures tend to be fragmented into individual
groups in HDC. 
Three galaxies in Virgo are bright enough to be in our survey; they are
assigned to a single group of 205 members in HDC, and a single group of 300
members in LDC.  Similarly, the four brightest galaxies in the Coma cluster
are in our survey.  They are all properly assigned to a single group in
both the HDC and LDC catalogs, containing 49 and 84 members, respectively.

Two measurements of the mass of each group are provided in
\citet{crooketal2007}, one based on the virial estimator and the other
based on the projected mass estimator.  The virial mass estimator is
computed from the line-of-sight velocity dispersion and mean harmonic
projected separation of group members.  The latter quantity is sensitive to
close pairs and can be noisy, in particular for groups not uniformly
sampled spatially.  The projected mass estimator \citep{heisleretal1985} is
designed to give equal weights to group members at all distances.  This
mass estimator depends on the mean eccentricity of the orbits and is
parameterized by an overall coefficient $f_{\rm pm}$ that typically is not
measured and must therefore be assumed.  The parameter $f_{\rm pm}$ ranges
from $32/\pi$ for isotropic orbits to $64/\pi$ for radial orbits,
independent of the mass distribution.  The \citet{crooketal2007} catalog
assumes $f_{\rm pm}=32/\pi$, which yields the smallest mass.

\begin{table}
\begin{center}
  \caption{Environment of MASSIVE galaxies}
  \begin{tabular}{llll}
   \hline
   Environment            &  HDC  & LDC  & 2M$++$ \\
   \hline
   Groupless             &   26  & 12   & 23  \\
   In groups              &   90  & 104  & 93 \\
   Brightest group galaxy &   65  & 70   & 70 \\
   \hline
  \end{tabular}
\end{center}
Notes.
Number of MASSIVE galaxies that are (1) group-less, i.e., ``isolated'' and have no group members;
(2) in groups of three or more members; and (3) the brightest galaxy in its group.
Three galaxy group catalogs (all based on 2MASS) are shown: the high-density-contrast (HDC) 
and low-density-contrast (HDC) catalogs of \citet{crooketal2007} and the 2M$++$ catalog \citep{lavauxhudson2011}.
\vspace{0.3in}
\label{table:group}
\end{table}

\begin{figure}
\vspace{-1.2in}
\hspace{-.2in}
\includegraphics[width=4.1in]{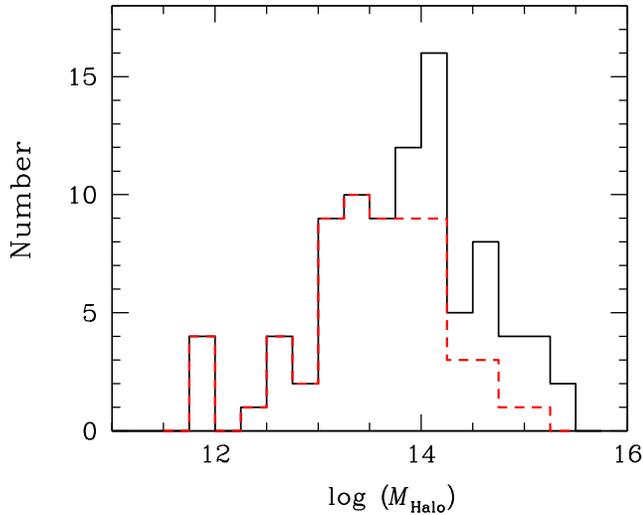}
 \vspace{-1.2in}
 \caption{Distribution of dark matter halo masses for the 90 MASSIVE
   galaxies that reside in groups in the HDC catalog (black histogram).
   The halo mass is obtained from the virial mass estimator
   (Sec~4.1). Among the 90, 65 are the brightest group galaxies (BGG) in
   their respective groups (red histogram).  The two histograms show that
   the 39 MASSIVE galaxies in the lower-mass groups are always the BGGs,
   whereas for those in groups with $M_{\rm halo} \ga 10^{13.75} \mdot$, 
   about 50\% are not BGGs. }
\label{group}
\vspace{0.3in}
\end{figure}

Table~2 lists the statistics of the environment of our candidate galaxies
classified by the HDC and LDC group catalogs.  As expected, more galaxies
are identified as being in groups in the LDC catalog.  Figure~\ref{group}
plots the distribution of the virial halo mass of the HDC groups in which
the MASSIVE galaxies reside.  The black histogram shows the halo
distribution of all 90 galaxies in groups, and the red histogram plots the
subset of 65 brightest group galaxies (BGGs).  The agreement of the two
histograms at $M_{\rm halo} \la 10^{13.75} \mdot$ indicates that the 39
MASSIVE galaxies in these lower-mass groups are all BGGs.  By contrast,
only 26 of the 51 galaxies in the higher-mass groups are BGGs.

\subsection{2M$++$ Group Catalog}

As a comparison study, we examine the environmental properties of the
MASSIVE galaxies in the 2M$++$ galaxy redshift catalog of
\citet{lavauxhudson2011}.  This more recent compilation of 69,160 galaxy
redshifts is based on the 2MASS photometric catalog for target selection
and uses primarily the redshifts from SDSS-DR7, 6dfGRS, and 2MRS.  The
catalog covers nearly the full sky and reaches depths of $K=12.5$, in
comparison to $K = 11.75$ mag for 44,599 galaxies in 2MRS.  Groups in this
catalog are identified by the FOF algorithm with linking parameters of 0.64
Mpc and $1000 \kms$. The corresponding overdensity threshold of 80 is the
same as the HDC catalog of Crook. The group list contains 4002 groups with
three or more members up to redshift distance of $20,000 \kms$.  The members of
the nearest two clusters Virgo and Fornax are not properly identified by
the FOF algorithm and are assigned manually.

\begin{figure}
\vspace{-1.2in}
\hspace{-.2in}
\includegraphics[width=4.1in]{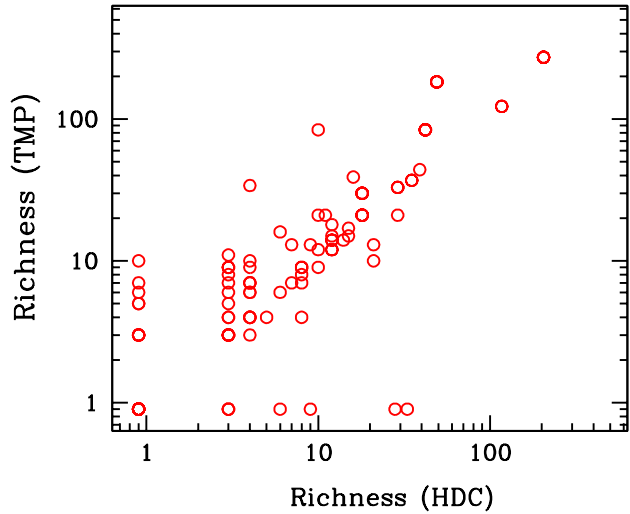}
 \vspace{-1.2in}
 \caption{Comparison of group membership in the 2MRS HDC catalog and 2M$++$
   catalog for MASSIVE galaxies.  The points with richness below 1 along
   each axis represent galaxies that are identified to reside in groups by
   only one catalog.  These are mostly lower-mass groups with a handful
   members.}
\label{group_compare}
\vspace{0.3in}
\end{figure}

A total of 93 MASSIVE galaxies are identified to reside in groups by the
2M$++$ catalog, similar to 90 in HDC (see Table~2).
Figure~\ref{group_compare} shows that the numbers of group members are
reasonably consistent between the two catalogs.
For the handful galaxies that are assigned to a group in one catalog but
not in the other, most of them reside in small groups of low velocity
dispersion and low membership, a regime that is more sensitive to the
different linking parameters and parent samples used in the two catalogs.

\subsection{Groupless Galaxies}

Not all MASSIVE galaxies are associated with groups (of three or more members)
in the Crook or 2M$++$ catalogs: 26 galaxies are not in groups according
HDC, and 23 galaxies are not in 2M$++$ groups.  Among these, 17 galaxies
are groupless in both catalogs.  These galaxies are relatively isolated
and presumably live in low-density environments.  Any satellite galaxy, if
present, is likely to be faint.

A handful of these 17 groupless galaxies had been targeted for X-ray
observations.  Three have archival Chandra and/or XMM-Newton observations:
NGC 57, NGC 4555, and NGC 7052.  The X-ray luminosity of the thermal
component in the $0.5–2$ keV band for the three galaxies are: $10^{41.19\pm
  0.02}$ erg s$^{-1}$ (NGC 57), $10^{41.27\pm 0.04}$ erg s$^{-1}$ (NGC
4555), and $10^{41.17+0.02-0.03}$ erg s$^{-1}$ (NGC 7052)
\citep{mulchaeyjeltema2010}.  The X-ray halos of NGC 57 and NGC 4555 both
have $kT \sim 0.9$ keV and extend to 50 to 60 kpc
\citep{osullivanponman2004, osullivanetal2007}.  
NGC 7052 has an X-ray halo of $kT \sim 0.48$ keV
\citep{memolaetal2009} and a central AGN with $L_x \sim 3\times 10^{40}$
erg s$^{-1}$ \citep{donatoetal2004}.

These groupless galaxies and other galaxies in low-richness groups in our
survey form an interesting subsample of targets for further studies. For
instance, we are investigating whether the groupless galaxies have faint
optical companions and satisfy the criterion of being fossil groups
\citep{ponmanetal1994, jonesetal2003}.  Overall, our survey galaxies span
only a factor of $\sim 3$ in stellar mass (Figure~\ref{mass_distance}) but a
much wider range in halo mass (Figure~\ref{group}) and group membership
(Figure~\ref{group_compare}), providing an excellent sample for studying
environmental effects on galaxy formation \citep{mulchaeyjeltema2010}.

\section{Observations}

\begin{figure*}
\vspace{-1.4in}
\hspace{-.5in}
\includegraphics[width=8.5in]{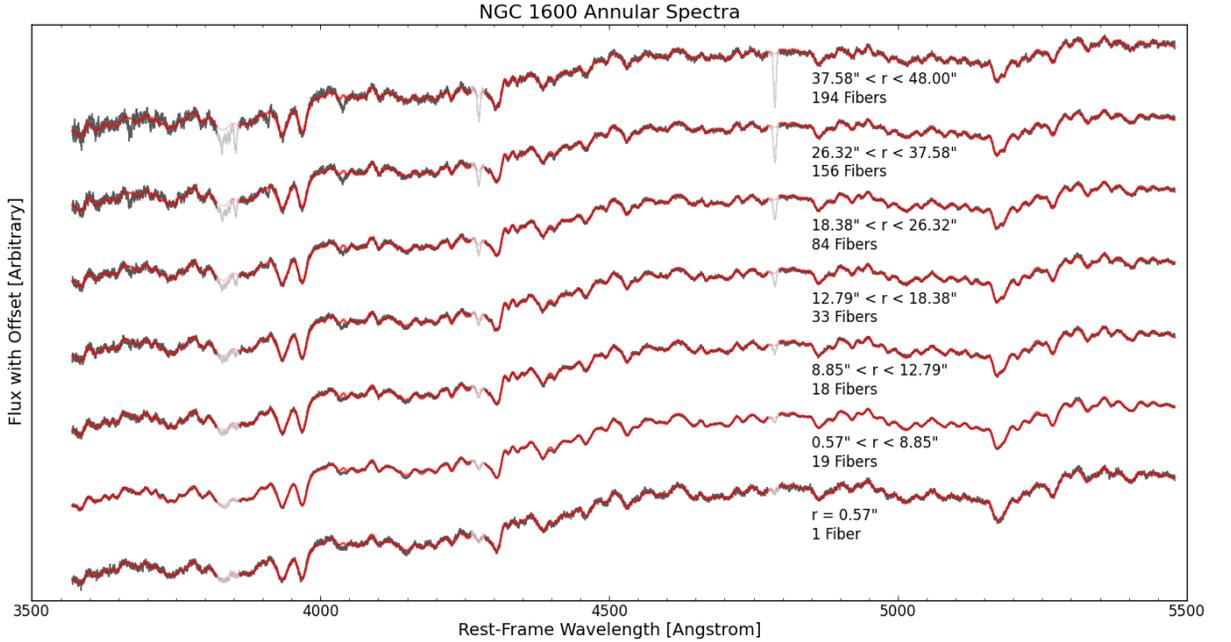}
\vspace{-1.4in}
\caption{Mitchell spectra for NGC~1600 in 7 radial bins. The bin
  size and number of fibers in each bin are labeled.  The best-fit
  spectra from pPXF are overlaid in red. }
\label{spectra}
\vspace{.3in}
\end{figure*}

\subsection{Large-format IFS}

Our large-scale IFS observations are performed with the Mitchell
Spectrograph \citep{hilletal2008a} on the 2.7 m Harlan J. Smith Telescope
at McDonald Observatory.  The Mitchell Spectrograph is an optical
integral-field spectrograph with a large field of view (107\arcsec $\times$
107\arcsec) and 4.1\arcsec\ diameter fibers.  The 246 fibers are
evenly-spaced and assembled in an array similar to Densepak on the WIYN
telescope \citep{bardenetal1998} with a one-third filling factor.

We use the low-resolution blue setting ($R \approx 850$) of the Mitchell
Spectrograph.  The wavelength coverage spans 3650 to 5850\AA, including
the Ca H+K region, the G-band region, H$\beta$, the Mg{\it b} region, and
several Fe absorption features. The spectral resolution varies spatially
and with wavelength but has an average of 5\AA\ FWHM, corresponding
to a dispersion of $\sim 1.1$\AA\ pixel$^{-1}$ and $\sigma \sim 100 \kms$ in the red part
of the spectrum to $\sim 150 \kms$ in the blue part.  

We observe each galaxy with three dither positions of equal exposure time
to obtain a contiguous coverage of the field of view.  For each dither
position, we interleave a ten-minute exposure on sky with two twenty-minute
on-target science frames.  Each galaxy is therefore observed for a total of
$\approx 2$ hours on source. With this observing strategy, we typically
reach S/N above 50 in the central fiber alone.  For the outer fibers, we
co-add the fibers and create spatial bins with a minimum of S/N$=20$ per
bin.  Our binning procedure provides a good combination of spatial
resolution and S/N, resulting in $\sim 30$ to 60 spatial bins per galaxy
and a median S/N from 25 to more than 30 for each of the ten galaxies we
have analyzed thus far.  Even in the outermost radial bin covered by the
IFU, our data have sufficient S/N to provide multiple angular bins (see
example in Figs.~\ref{maps} and \ref{lick_maps}).  Comparable S/N
requirements are used for dynamical orbit modeling of luminous and dark
matter in Coma galaxies \citep{thomasetal2007} and for \mbh\ measurements
using combined wide-field and AO IFU data (e.g. \citealt{ruslietal2013,
  mcconnelletal2012}).
With these data, we expect to constrain stellar population gradients larger
than roughly a tenth of a dex per decade in radius (e.g.,
\citealt{greeneetal2013}).  In a handful of galaxies, we have integrated
substantially longer (i.e., 6-8 hours) on an off-nucleus pointing (e.g.,
\citealt{murphyetal2011}).  With these cases we are able to roughly double
our radial coverage at comparable S/N.

The data reduction is performed using the Vaccine package
\citep{adamsetal2011, murphyetal2011}.  Flux calibration and final
reduction are done with the software developed for the VENGA project
\citep{blancetal2009,blancetal2013}.  The flux calibration is quite robust,
with $<10\%$ disagreement in continuum shape between the central Mitchell
fiber and SDSS spectra when available
\citep{greeneetal2012,greeneetal2013}.
Figure~\ref{spectra} shows the spectra for a range of radial bins for NGC~1600
from our Mitchell IFS data taken in October 2013. 
The resulting stellar kinematics and stellar populations for NGC~1600 are presented
in Sec.~6 below.

In 2010 we conducted a precursor study to MASSIVE with the Mitchell
spectrograph \citep{greeneetal2012}.  Galaxies were selected to have red
colors ($u-r>2.2$) and velocity dispersions $> 150 \kms$ (as measured by
the SDSS) within the redshift range $0.01 < z < 0.02$. The galaxies span a
mass range of $M^* \approx 10^{10.4-11.5}\msun$, less massive than MASSIVE
galaxies but still generally more massive than the \atlas\ galaxies.  Fifty
galaxies have been observed to date, and we analyzed their stellar
population properties in \citet{greeneetal2013} and their kinematic
properties in \citet{raskuttietal2014}.  Because they were observed in an
identical manner to the MASSIVE galaxies, we will use them as a
complementary sample for comparison studies.

\subsection{AO-assisted IFS}

For a subset of MASSIVE galaxies suitable for AO-assisted observations, we
are acquiring high-resolution data to perform new measurements of black
hole masses \mbh\ using stellar dynamics.  We use NIFS and the ALTAIR
adaptive optics system with both natural guide star (NGS) and laser guide
star (LGS) on the 8 m Gemini Observatory North telescope, and OSIRIS
\citep{larkinetal2006} and LGS-AO system on the 10 m W. M. Keck I
telescope.

The literature contains \mbh\ measurements for seven galaxies in our
MASSIVE sample (see Sec~\ref{sec:SMBH}).  We are preparing \mbh\
measurements for six additional MASSIVE galaxies based on our existing AO
data; seeing-limited IFS data at $\sim 0.4''$ resolution may yield up to
four more.  Our ongoing campaign with AO instruments will extend the sample
of \mbh\ in MASSIVE galaxies still further.

Because MASSIVE galaxies are selected for extreme stellar masses, and
because they mostly lie within a factor of two in distance (54-108 Mpc),
much of the variation in the angular sizes of influence of the central
black holes results from the cosmic scatter in \mbh.  In this case the
primary limiting factor for AO selection is central surface brightness:
fainter than $\mu_K = 13.5$ mag arcsec$^{-2}$, high-resolution observations
are prohibitively expensive.  Within the acceptable range of surface
brightnesses, we typically require four to eight hours of AO observations
per target, including science and sky frames, calibration stars, and
overheads.

\subsection{Deep $K-$band Imaging}
\label{sec:deepK}

Most of our science goals require a deep luminosity profile for each
galaxy.  Since 2MASS is shallow, we are obtaining deeper $K$-band imaging
for the MASSIVE sample using a combination of WFCAM on UKIRT and WIRCam on
CFHT.  We choose $K$-band because it (i) traces the old populations that
compose most of the stellar mass in early-type galaxies; (ii) minimizes
dust extinction; (iii) allows for uniform calibration using 2MASS; and (iv)
facilitates comparison of black hole masses across galaxy populations via
the \mbh\ -$L_K$ relation.

In order to trace the extended halos of luminous early-type galaxies and
measure accurate total magnitudes, it is desirable to reach a surface
brightness limit $\sim 3$ mag arcsec$^{-2}$ fainter than 2MASS (cf.\
Appendix~B of \citealt{laueretal2007}).  The 2MASS $3\sigma$ $K$ surface
brightness limit is $\mu_K = 18.6$ mag arcsec$^{-2}$
\citep{jarrettetal2000}, which corresponds roughly to the often quoted
$1\sigma$ value of $\mu_K \sim 20$ mag arcsec$^{-2}$. Thus, in terms of AB
mag, we are aiming to achieve a $3\sigma$ surface brightness limit of $\sim
23.6$ mag arcsec$^{-2}$ (3 mag in depth plus 2 mag AB conversion).

\section{Examples of Survey Science and Early Results}

The MASSIVE survey is designed to study the most massive galaxies in the
universe today, a parameter space that has not been systematically explored
with IFS to date.  With the nearly 2\arcmin\ field of view of the Mitchell
spectrograph, we cover about twice the effective radius of most galaxies in
the survey.  The additional AO data for a subset of the galaxies will probe
sub-arcsec scales down to the gravitational sphere of influence of the
central supermassive black hole ($\sim 100$ pc).  In the following sections
we discuss key science results that can be expected from the survey and
present some early results.  This list is by no means exhaustive.

\subsection{Stellar Mass-to-Light Ratio and IMF}

Our kinematic measurements at large radius, combined with Schwarzschild
orbit modeling, allow us to measure the dark matter halo mass and the
dynamically inferred stellar mass-to-light ratio \mldyn.  At the same time,
stellar population synthesis modeling of our Mitchell spectra in the blue
combined with our $K$-band imaging and space-based photometry in the
mid-infrared provide an independent measurement of the stellar mass,
yielding \mlpop.

The observed increase in the ratio of \mldyn/\mlpop\ in galaxies with
increasing \sig\ has been interpreted as a change in the IMF
\citep[e.g.,][]{treuetal2010,augeretal2010,cappellarietal2012,
  sonnenfeldetal2012,tortoraetal2013,duttonetal2013,
  barnabeetal2011,barnabeetal2013}, but it could also indicate a degeneracy
with the dark matter distribution
\citep[e.g.][]{thomasetal2011,wegneretal2012}.  A number of independent
approaches, including direct measurements of gravity-sensitive stellar
features and gravitational lensing, have pointed towards an IMF that
becomes more top-heavy in galaxies with higher stellar velocity dispersions
\citep[e.g.,][]{conroyvandokkum2012mod, spinielloetal2014,ogurietal2014}.
Some recent results, however, are not consistent with an increasingly
top-heavy IMF in all systems \citep[e.g.,][]{smithlucey2013,ruslietal2013,
  smith2014}.  The sample size and dynamic range in mass of the MASSIVE
survey will improve the constraints on any possible mass dependence of the
IMF.

\begin{figure*}
\vspace{-.2in}
\hspace{.7in}
\includegraphics[width=6.in]{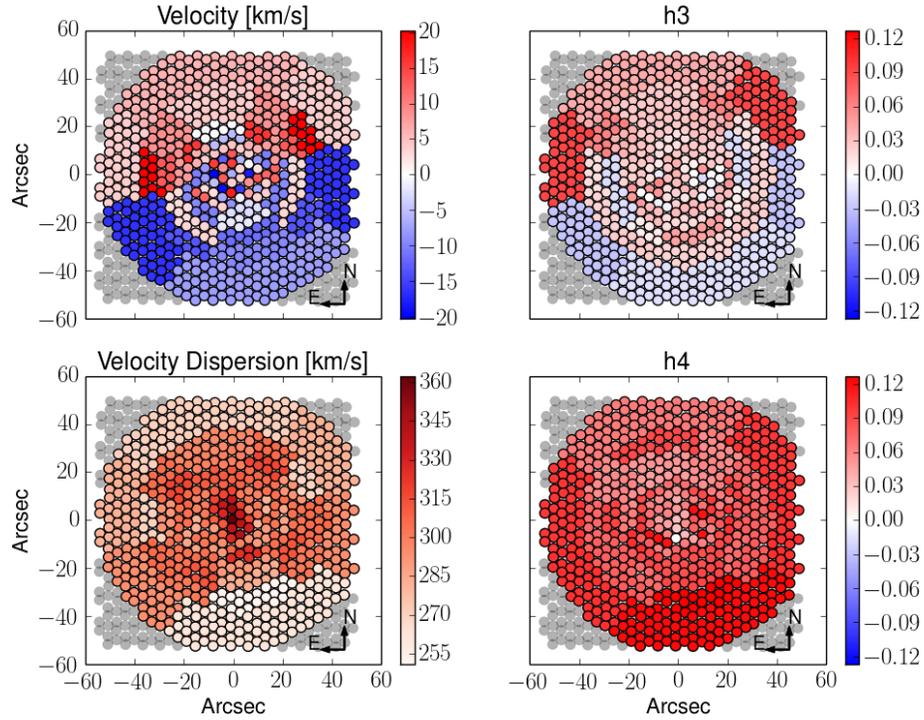}
\vspace{-.1in}
\caption{Kinematic maps of the line-of-sight velocity moments for NGC~1600.
  The four panels show the four Gauss-Hermite moments, $V, \sigma, h_3$, and $h_4$, respectively.
  The circles in each panel indicate the Mitchell IFS fibers.
  The fibers are grouped into spatial bins to ensure a minimum S/N of 20;
  the median S/N of the bins is 30.6 for NGC~1600.  Individual fibers near
  the center have S/N$\sim 60$.  The median errors over the spatial bins
  for the four moments are $17 \kms$, $24 \kms$, 0.051, and 0.057,
  respectively.}
\label{maps}
\end{figure*}

\begin{figure*}
\vspace{-1.5in}
\hspace{-.2in}
\includegraphics[width=7.5in]{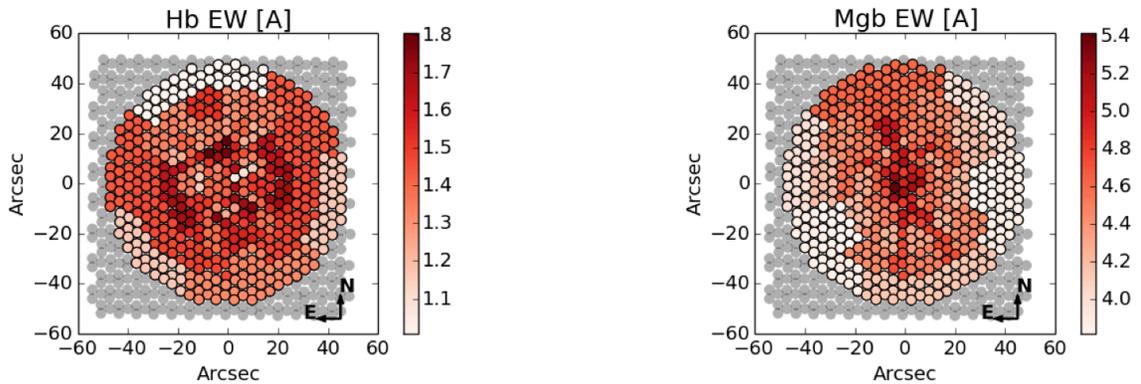}
\vspace{-1.5in}
\caption{Two-dimensional maps of the equivalent widths (in Angstroms) of
  the H$\beta$ (left) and Mg{\it b} (right) absorption lines in NGC~1600.
  Our Mitchell data have sufficient S/N to provide measurements in several
  angular bins at each radius.  The typical errors are 0.2\AA\ to 0.3\AA. }
\label{lick_maps}
\end{figure*}

As a demonstration of early results from our survey, we show in
Figure~\ref{maps} the 2-dimensional stellar kinematic maps for NGC~1600.
We use the penalized pixel-fitting (pPXF) method
\citep{cappellariemsellem2004} to extract the stellar line-of-sight
velocity distribution (LOSVD) function $f(v)$ from the absorption line
features in our spectra.  As input templates, we use the MILES library of
985 stellar spectra, covering the wavelength range of 3525-7500\AA\ at
2.5\AA\ (FWHM) spectral resolution \citep{sanchezetal2006}.  The pPXF
routine convolves the MILES stellar templates with $f(v)$ modeled as a
Gauss-Hermite series (up to order 6):
\begin{equation}
  f(v) \propto \frac{1}{\sqrt{2\pi \sigma^2}} e^{-\frac{(v-V)^2}{\sigma^2}}
     \left[ 1 + \sum_{m=3}^n h_m H_m \left( \frac{v-V}{\sigma}\right) \right] \,,
\label{gh}
\end{equation}
where $H_m(x)$ is the $m$th Hermite polynomial and given by 
\begin{equation}
   H_m(x)= \frac{1}{\sqrt{m!}}\, e^{x^2} \left( -\frac{1}{\sqrt{2}} \frac{\partial}{\partial x} \right)^m e^{-x^2} \, .
\end{equation}

Figure~\ref{maps} shows the 2-dimensional maps of the best-fit
Gauss-Hermite velocity moments $V, \sigma, h_3,$ and $h_4$ from our
Mitchell IFS observations of NGC~1600. The galaxy rotates slowly with $V
\la 20 \kms$ about its photometric minor axis.  The velocity dispersion
peaks at $360 \kms$ in the central fiber and declines radially by $\sim
20$\% out to $\sim 50\arcsec$. The flux-weighted $V/\sigma$ is $0.03 \pm
0.01$ for NGC~1600. Only two galaxies have such low $V/\sigma$ in \atlas
\citep{emsellemetal2011}.  We will discuss in separate papers the details
of this analysis, results for \mldyn, and tests of systematics including
spectral regions used in the fits and robustness of the higher-order
Gauss-Hermite moments (J. Thomas et al. 2014, in prep.; R. Janish et al. 2014, in
prep.).

\subsection{Radial Gradients and Assembly History}

Massive early-type galaxies have experienced dramatic size evolution, by
factors of 2-4, from $z \approx 2$ to the present
\citep[e.g.,][]{vandokkumetal2008}.  One way to understand the physical
mechanisms responsible for this growth is to study spatial gradients in the
stellar populations and kinematics beyond the half-light radius of
present-day ellipticals.  Since the dynamical times in the outskirts of
these galaxies are long, the stars can potentially remember their origin
both in their overall distribution \citep{naabetal2007,oseretal2010,
  hilzetal2013} and their degree of angular momentum
\citep[e.g.,][]{daviesetal1983,franxetal1991,krajnovicetal2011,
  wuetal2014,arnoldetal2014,naabetal2014,
  krajnovicetal2013,raskuttietal2014}.
 
Sensitive spectroscopic observations of stellar populations at large radius
are still relatively scarce
\citep[e.g.,][]{carollodanziger1994,mehlertetal2003,kelsonetal2006,
  weijmansetal2009,spolaoretal2010,puetal2010,
  pastorelloetal2014,martinnavarroetal2014,greeneetal2012,greeneetal2013}.
The MASSIVE survey will contribute the largest set of IFS data to date for
slowly rotating nearby early-type galaxies in a wide range of large-scale
environments. In combination with our deep $K$-band imaging, we will also
investigate any correlations between rotation as a function of radius and
isophotal shape \citep[e.g.,][]{benderetal1989,arnoldetal2014}.

Figure~\ref{lick_maps} shows the 2-dimensional maps of the equivalent
widths (EW) of the H$\beta$ (left) and Mg{\it b} (right) absorption
features for NGC~1600.  We measure the standard Lick indices
\citep{faberetal1985, wortheyetal1994} using the IDL code {\it lick\_ew}
\citep{gravesschiavon2008}.  In \citet{greeneetal2012} we demonstrated that
because these indices are defined for flux-calibrated spectra, our
measurements are on a standard system without any additional offsets.  Very
low levels of emission-line infill can often contaminate our EW
measurements, particularly of H$\beta$ (e.g., \citealt{gravesetal2007}).
We perform an iterative fit \citep{greeneetal2013} to the H$\beta$+[O
{\small III}] region, using the stacked spectra from \citet{gravesetal2009}
as templates.  Our uncertainties are dominated by sky subtraction, so we
scale our fiducial sky model by up to $\pm 5$\% and repeat our full
procedure to determine the uncertainties.  Given our typical S/N ratios of
30 - 100 at $\sim 2 R_e$, we achieve S/N of 15-80 in the H$\beta$ and
Mg{\it b} line indices at the outer edge of the Mitchell IFU.

Figure~\ref{lick_maps} shows a characteristic radial decline in Mg{\it b}
EW, mostly due to the well-known decline in metallicity with radius in
these early-type galaxies.  In contrast, the H$\beta$ EW is relatively flat
with radius, with perhaps a subtle trend of falling at the outer parts,
reflecting the uniform old age of this galaxy.

\subsection{Black Hole-Galaxy Correlations}

New kinematic data and modeling efforts in the past several years have
substantially expanded and revised dynamical measurements of \mbh.  As
samples of dynamical black hole masses increase at both the highest
masses \citep[e.g.,][]{mcconnelletal2011a,mcconnelletal2011b,
  mcconnelletal2012, ruslietal2011, ruslietal2013, vdboschetal2012,
  walshetal2013} and in spiral galaxies
\citep[e.g.,][]{greeneetal2010,kuoetal2011,beifiorietal2012,
  sunetal2013}, it becomes increasingly clear that more data are
needed to better quantify the intrinsic scatter and mass dependence in
the scaling relations between \mbh\ and properties of their host
galaxies \citep{mcconnellma2013,kormendyho2013}.

A systematic survey of dynamical black hole masses in the most massive
galaxies (without preselection based on current scaling relations) will
substantially improve our leverage on the intrinsic scatter in the
relations as a function of mass, which may discriminate between different
models for galaxy-black hole coevolution
\citep[e.g.,][]{peng2007,hirschmannetal2010,jahnkemaccio2011,AA13}.
Knowledge of the intrinsic scatter in \mbh\ is crucial for calculating the
quiescent black hole mass function, as is understanding whether stellar
mass or stellar velocity dispersion is a better predictor of black hole
mass \citep[e.g.,][]{laueretal2007,laueretal2007scatter}.  In addition to
providing key constraints on current theories of black hole and galaxy
growth, these scaling relations are also a critical input in numerous
applications that rely on black hole demographics, e.g., the predicted
contributions from merging supermassive black hole binaries to the
gravitational wave background targeted by the ongoing pulsar timing
experiments \citep{Haasteren11, Demorest13, shannonetal2013} and LISA.

\begin{figure}
\vspace{.2in}
\includegraphics[width=3.5in]{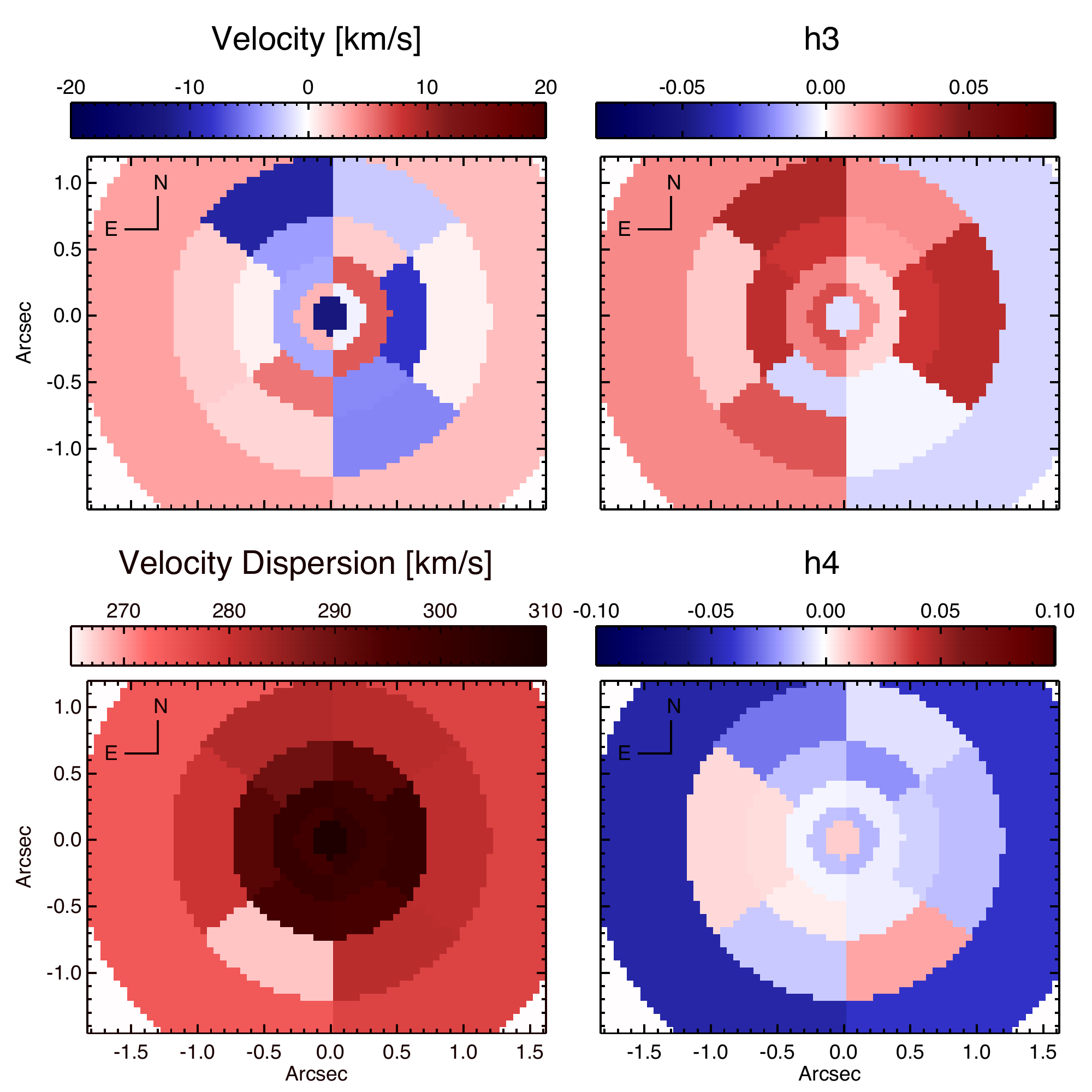}
\vspace{.2in}
\caption{Kinematic maps of the line-of-sight velocity moments for the inner
  $3\arcsec\times 2\arcsec$ region of NGC~5777 from Gemini NIFS.  The
  moments are defined the same way as in Figure~\ref{maps}.  The median
  errors are $4.6 \kms$ in $V$, $5.4 \kms$ in $\sigma$, 0.016 in $h_3$, and
  0.015 in $h_4$.}
\label{nifs}
\vspace{0.3in}
\end{figure}


Cores in central light profiles within a few ∼100 pc are often seen in
massive galaxies. These cores are thought to be a sign of black hole
scouring \citep{begelmanetal1980}, consistent with several scaling
relations between core size and other galaxy properties (e.g.,
\citealt{faberetal1997, ferrarese2006, laueretal2007, kormendybender2009,
  ruslietal2013core}).  As part of our dynamical modeling, we see evidence
for an excess of tangential orbits at the galaxy centers, consistent with
black hole scouring.  We will use our orbit modeling to investigate the
connection between the central black hole mass and nuclear galaxy structure.


Figure~\ref{nifs} shows two-dimensional stellar kinematics for the central
region of NGC~5557 from our observations with NIFS and LGS-AO on Gemini
North.
We use stellar templates from Gemini's NIFS/GNIRS template library and fit
LOSVD-convolved templates to the $K$-band CO bandhead features in each
galaxy spectrum.
Figure~\ref{nifs} shows the resulting four Gauss-Hermite velocity moments
as defined in Equation~(\ref{gh}).  When applied to stellar orbit, our combined
NIFS and Mitchell data yield a black hole mass of $3.9^{+1.0}_{-1.3} \times
10^9 \msun$ in NGC 5557 (McConnell et al. 2014, in prep.).

\subsection{The $R_e - L$ and $\sigma - L$ Relations}

The tight scaling relations among size, luminosity, and stellar velocity
dispersion of early-type galaxies
\citep[e.g.,][]{faberjackson1976,kormendy1977, dressleretal1987} have long
been used to constrain galaxy assembly \citep[e.g.,][and references
therein]{boylankolchinma2005,robertsonetal2006}.  With our spatially
resolved stellar kinematics and deep $K-$band imaging, we will refine the
measurements of the galaxy scaling relations by adding galaxies at the most
massive end \citep[e.g.,][]{bernardietal2003fp}.

\begin{figure}
\vspace{-0.8in}
\hspace{-.2in}
\includegraphics[width=4.1in]{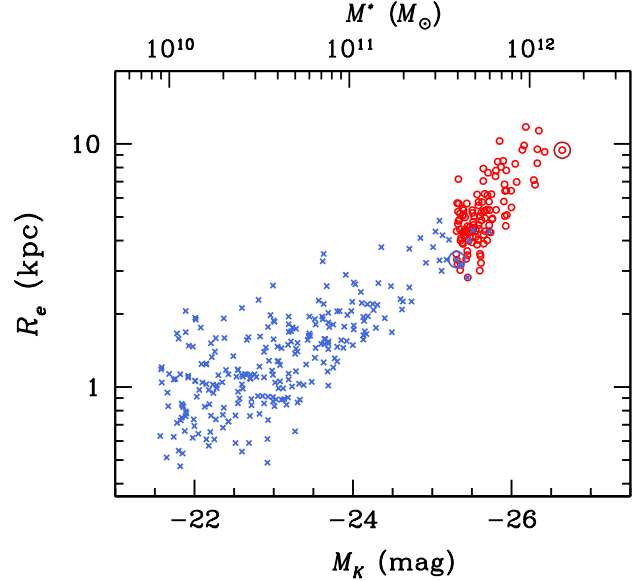}
\vspace{-1.2in}
\caption{2MASS effective radius versus absolute $K$-band magnitude for galaxies in
  the MASSIVE (red circles) and \atlas surveys (blue crosses). 
  The big circles indicate NGC 4889 (red) and M87 (blue). }
\vspace{0.3in}
\label{size_mass}
\end{figure}

\begin{figure}
\vspace{-1.2in}
\hspace{-.2in}
\includegraphics[width=4.1in]{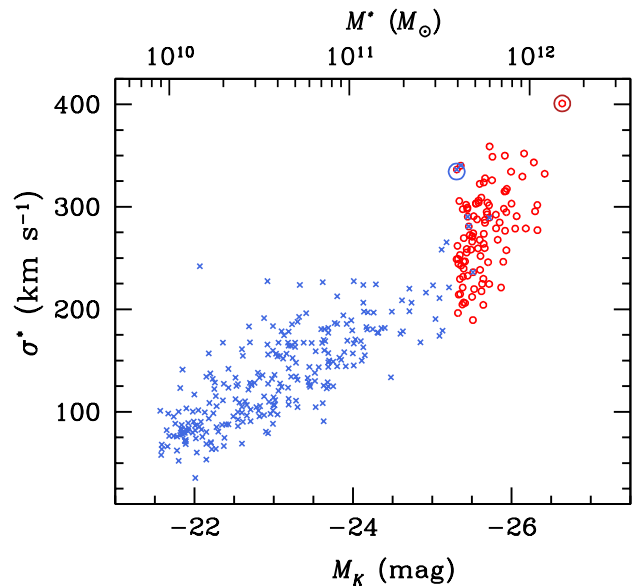}
 \vspace{-1.2in}
 \caption{Stellar velocity dispersion versus absolute $K$-band magnitude for
   galaxies in the MASSIVE survey (red circles) and \atlas survey (blue
   crosses).  A total of 103 MASSIVE galaxies have measured $\sigma$ in
   HyperLeda and/or NSA.  The big circles indicate NGC 4889 (red) and M87
   (blue).  }
\vspace{0.3in}
\label{sigma_mass}
\end{figure}

Figure~\ref{size_mass} plots the 2MASS $R_e$ and $M_K$ relation
\citep[e.g.,][]{kormendy1977} for MASSIVE and \atlas galaxies.
Figure~\ref{sigma_mass} shows the stellar velocity dispersion and $M_K$ for
the 103 MASSIVE galaxies with existing $\sigma$ measurements.  We emphasize
that no cuts are made on either $R_e$ or $\sigma$ in our sample selection.
These plots are only meant to illustrate the demographics of our survey
galaxies based on currently available data.  We will improve these
measurements to address possible biases in 2MASS $M_K$ (Sec~\ref{sec:K})
and massive galaxy sizes (e.g., \citealt{bernardietal2014RL}) and to study
the distribution of our galaxies in projections of the fundamental plane
\citep[e.g.,][]{laueretal2007, kormendyetal2009}.

\subsection{X-Ray Gas and Halo Mass}

Roughly 30\% of the galaxies in the MASSIVE survey have archival
Chandra/XMM X-ray observations that are sensitive enough to detect
thermal emission from the hot halo gas.  If the gas is in thermal
equilibrium, then the ratio $L_X$/$M^*$ reflects the ratio of dark
matter halo to stellar mass.  Empirically, large scatter (factor of
$\sim 100$) is found between $L_X$ and $L_K$
\citep[e.g.,][]{formanetal1985, fabbiano1989}, and there are hints
that the slope and scatter depend on environment
\citep{mulchaeyjeltema2010}.  Likely at play are both intrinsic
scatter in the relation between stellar and dark halo mass, and
non-equilibrium conditions in the hot gas, e.g., due to AGN
feedback \citep{diehlstatler2008, dunnetal2010}.

There is a hint of a tighter correlation between $L_X$ and total dynamical
mass (stellar and dark matter halo) than $L_K$
\citep[][]{mathewsetal2006,kimfabbiano2013}, but the sample with
independent dynamical halo masses and deep X-ray observations is small, and
M87 is still the most massive galaxy included. With MASSIVE, we will
revisit the $L_X/L_K$ and $L_X/M_{\rm tot}$ relations for a large and
well-defined sample with uniform dynamical halo masses and a range of
environments. We may also explore the importance of radio jets in keeping
the halo gas from cooling \citep[][]{allenetal2006,mcnamaranulsen2007}.

\section{Summary}

MASSIVE is a comprehensive IFS survey of a volume-limited and mass-selected
sample of the most massive early-type galaxies within $\sim 108$ Mpc.
MASSIVE is the first IFS survey to specifically target galaxies with
$M^*>10^{11.5}$~\msun.  We exploit the large (107\arcsec$\times$107\arcsec)
areal coverage of the Mitchell Spectrograph to obtain stellar population
and kinematic information beyond twice the effective radius of our
galaxies, while using AO-assisted IFS data on small scales to probe the
sphere of influence of the supermassive black hole.  The sample galaxies
span a narrow range in stellar mass, but a wide range in stellar velocity
dispersion, size, halo mass and large-scale environment. Thus, we are
poised to determine the relationships between central black hole mass,
stellar mass, and dark halo mass for the most massive galaxies in the
universe today.

\acknowledgements

We thank Joshua Adams, Akos Bogdan, Stephen Chen, Bill Forman, Jim Gunn,
and Christine Jones for useful discussions.  This survey is supported in
part by NSF AST-1411945 and AST-1411642.  C.-P.M. is supported in part by
grants from the Simons Foundation (No.~224959) and NSF AST-1009663.
N.J.M. is supported by the Beatrice Watson Parrent Fellowship.  J.D.M. is
supported by an NSF Astronomy and Astrophysics Postdoctoral Fellowship
(AST-1203057).  We thank the Cynthia and George Mitchell Foundation for
funding the Mitchell Spectrograph, and Gary J. Hill and Phillip MacQueen on
their continued work to make the Mitchell Spectrograph a successful
instrument.  Dave Doss, Kevin Meyer, Brian Roman, John Kuehne, Coyne Gibson
and all of the staff at McDonald Observatory have helped immensely with
collection of these data.

This research has made use of the HyperLeda database and the NASA/IPAC
Extragalactic Database (NED) which is operated by the Jet Propulsion
Laboratory, California Institute of Technology, under contract with the
National Aeronautics and Space Administration.


\appendix
\makeatletter
\def\@seccntformat#1{Appendix\ \csname the#1\endcsname\quad}
\makeatother


\begin{table*}
  \caption{116 candidate MASSIVE galaxies}
  \label{big_table}
  \begin{center}
  \begin{tabular}{rccccccccccll}
   \hline
   Galaxy &  R.A. &  Dec. & $D$   &  $K$  & $A_V$ & $M_K$  & $\sigma^{\rm HL}$ & $\sigma^{\rm NSA}$ & $R_e^{\rm 2MASS}$ & $R_e^{\rm NSA}$ & Env. & Note  \\
          & (deg) & (deg) & (Mpc) & (mag) & (mag) &  (mag) & (km/s)           & (km/s)           & (arcsec)        & (arcsec)       &      &  \\
    (1)   &  (2)  &  (3)  &  (4)  &  (5)  &  (6)  &   (7)  &  (8)             &  (9)             &    (10)         &  (11)          & (12) & (13) \\
   \hline
   \hline

  NGC 0057 &     3.8787 &    17.3284 &    76.3 &  8.68 &   0.212 & $-$25.75 &  326 &  &  13.2 &    27.0 &   1  &   \\
  NGC 0080 &     5.2952 &    22.3572 &    81.9 &  8.92 &   0.168 & $-$25.66 &  260 &  &  15.7 &    32.2 &  14 B &   \\
  NGC 0128 &     7.3128 &     2.8641 &    59.3 &  8.52 &   0.079 & $-$25.35 &  215 &  &  10.5 &    18.0 &   1  &   \\
  NGC 0227 &    10.6534 &  $-$1.5288 &    75.9 &  9.09 &   0.084 & $-$25.32 &  262 &  &   8.7 &    27.2 &   4 B &   \\
  NGC 0315 &    14.4538 &    30.3524 &    70.3 &  7.96 &   0.177 & $-$26.30 &  296 &  &  20.0 &    25.1 &   6 B &   \\
  NGC 0383 &    16.8540 &    32.4126 &    71.3 &  8.48 &   0.194 & $-$25.81 &  279 &  &  15.5 &    20.5 &  29  & $=$4ZW038 \\
  NGC 0393 &    17.1540 &    39.6443 &    85.7 &  9.23 &   0.120 & $-$25.44 &  233 &  &  11.0 &         &   1  &   \\
  NGC 0410 &    17.7453 &    33.1520 &    71.3 &  8.38 &   0.161 & $-$25.90 &  298 &  &  16.8 &    31.6 &  29 B &   \\
  NGC 0467 &    19.7922 &     3.3008 &    75.8 &  9.01 &   0.092 & $-$25.40 &  247 &  &  14.5 &    21.5 &   1  &   \\
PGC 004829 &   20.1287 &    50.1445 &    99.0 &  9.74 &   0.554 & $-$25.30 &       &  &   7.3 &         &   1  &   \\

  NGC 0499 &    20.7978 &    33.4601 &    69.8 &  8.74 &   0.193 & $-$25.50 &  266 &  &  11.6 &    15.6 &  35  & \\
  NGC 0507 &    20.9164 &    33.2561 &    69.8 &  8.30 &   0.170 & $-$25.93 &  295 &  &  23.0 &    38.4 &  35 B & \\
  NGC 0533 &    21.3808 &     1.7590 &    77.9 &  8.42 &   0.084 & $-$26.05 &  279 &  &  21.9 &    40.7 &   3 B & \\
  NGC 0545$^\dagger$ &21.4963 &  $-$1.3402 & 74.0 &     &   0.114 &          &  250 &  &       &         &  32 B & A194\\
  NGC 0547 &    21.5024 &  $-$1.3451 &    74.0 &  8.49 &   0.113 & $-$25.83 &  262 &  &  25.1 &    19.7 &  32   & A194\\
  NGC 0665 &    26.2338 &    10.4230 &    74.6 &  8.88 &   0.242 & $-$25.51 &  190 &  &  11.5 &    13.7 &   4 B & \\
 UGC 01332 &    28.0755 &    48.0878 &    99.2 &  9.48 &   0.557 & $-$25.57 &      &  &  12.9 &         &   8 B & \\
  NGC 0708 &    28.1937 &    36.1518 &    69.0 &  8.57 &   0.247 & $-$25.65 &  230 &  &  23.7 &         &  39 B & A262 \\
 UGC 01389 &    28.8778 &    47.9550 &    99.2 &  9.63 &   0.519 & $-$25.41 &      &  &   9.2 &         &    8 & \\
  NGC 0741 &    29.0874 &     5.6289 &    73.9 &  8.30 &   0.144 & $-$26.06 &  291 &  &  19.5 &    26.9 &   5 B & \\

  NGC 0777 &    30.0622 &    31.4294 &    72.2 &  8.37 &   0.128 & $-$25.94 &  318 &  &  14.6 &    18.6 &   7 B & \\
  NGC 0890 &    35.5042 &    33.2661 &    55.6 &  8.25 &   0.212 & $-$25.50 &  212 &  &  16.7 &         &    1 & \\
  NGC 0910 &    36.3616 &    41.8243 &    79.8 &  9.20 &   0.157 & $-$25.33 &  249 &  &  13.6 &         &   29 & A347 \\
  NGC 0997 &    39.3103 &     7.3056 &    90.4 &  9.42 &   0.380 & $-$25.40 &      & &    9.6 &    23.5 &   3 B & \\
  NGC 1016 &    39.5815 &     2.1193 &    95.2 &  8.58 &   0.085 & $-$26.33 &  302 &  &  18.1 &    26.8 &   8 B & \\
  NGC 1060 &    40.8127 &    32.4250 &    67.4 &  8.20 &   0.532 & $-$26.00 &  303 &  &  16.8 &    36.9 &  12 B & \\
  NGC 1066 &    40.9579 &    32.4749 &    67.4 &  8.89 &   0.563 & $-$25.31 &      &  &  17.5 &    26.6 &  12  & \\
  NGC 1132 &    43.2159 &  $-$1.2747 &    97.6 &  9.26 &   0.176 & $-$25.70 &  246 &  &  16.1 &    30.9 &   3 B & \\
  NGC 1129 &    43.6141 &    41.5796 &    73.9 &  8.24 &   0.309 & $-$26.14 &  330 &  &  26.4 &    30.2 &  33 B & \\
  NGC 1167 &    45.4265 &    35.2056 &    70.2 &  8.64 &   0.496 & $-$25.64 &  204 &  &  20.7 &    29.7 &   3 B & \\

  NGC 1226 &    47.7723 &    35.3868 &    85.7 &  9.21 &   0.526 & $-$25.51 &  271 &  &  12.5 &         &   3 B & \\
   IC 0310 &    49.1792 &    41.3248 &    77.5 &  9.15 &   0.445 & $-$25.35 &  230 &239 & 11.8 &   15.3 &  117 & Perseus/A426 \\
  NGC 1272 &    49.8387 &    41.4906 &    77.5 &  8.69 &   0.441 & $-$25.80 &  292 &   & 20.7 &    31.5 &  117 & Perseus/A426 \\
 UGC 02783 &    53.5766 &    39.3568 &    85.8 &  9.27 &   0.447 & $-$25.44 &  299 &  &   8.4 &     9.0 &   4 B & \\
  NGC 1453 &    56.6136 &  $-$3.9688 &    56.4 &  8.12 &   0.289 & $-$25.67 &  328 &  &  16.0 &         &  12 B & \\
  NGC 1497 &    60.5283 &    23.1329 &    87.8 &  9.48 &   0.602 & $-$25.31 &  249 &  &  10.3 &         &    1 & \\
  NGC 1600 &    67.9161 &  $-$5.0861 &    63.8 &  8.04 &   0.118 & $-$25.99 &  334 &  &  20.8 &         &  16 B & \\
  NGC 1573 &    68.7666 &    73.2624 &    65.0 &  8.56 &   0.377 & $-$25.55 &  303 &  &  13.9 &         &  15 B & \\
  NGC 1684 &    73.1298 &  $-$3.1061 &    63.5 &  8.69 &   0.159 & $-$25.34 &  306 &  &  15.8 &         &  11 B & \\
  NGC 1700 &    74.2347 &  $-$4.8658 &    54.4 &  8.09 &   0.119 & $-$25.60 &  239 & &   13.4 &         &   4 B & \\
  
  NGC 2208 &    95.6444 &    51.9095 &    84.1 &  9.04 &   0.408 & $-$25.63 &  225 &  &  14.2 &         &    1 & \\
  NGC 2256 &   101.8082 &    74.2365 &    79.4 &  8.67 &   0.359 & $-$25.87 &  221 &  &  20.9 &         &  10 B & \\
  NGC 2274 &   101.8224 &    33.5672 &    73.8 &  8.68 &   0.286 & $-$25.69 &  295 &  &  15.0 &         &   6 B & \\
  NGC 2258 &   101.9425 &    74.4818 &    59.0 &  8.23 &   0.351 & $-$25.66 &  287 &  &  18.6 &         &   3 B & \\
  NGC 2320 &   106.4251 &    50.5811 &    89.4 &  8.85 &   0.189 & $-$25.93 &  315 &  &  10.6 &         &  18 B & \\
 UGC 03683 &   107.0582 &    46.1159 &    85.1 &  9.16 &   0.253 & $-$25.52 &  291 &  &  11.2 &         &   4 B & \\
  NGC 2332 &   107.3924 &    50.1823 &    89.4 &  9.40 &   0.241 & $-$25.39 &  232 &  &   8.9 &         &  18  & \\
  NGC 2340 &   107.7950 &    50.1747 &    89.4 &  8.88 &   0.203 & $-$25.90 &  246 &  &  19.7 &         &  18  & \\
 UGC 03894 &   113.2695 &    65.0791 &    97.2 &  9.37 &   0.175 & $-$25.58 &  304 &  &  12.2 &    17.8 &   4 B & \\
  NGC 2418 &   114.1563 &    17.8839 &    74.1 &  8.95 &   0.102 & $-$25.42 &  247 &  &  11.7 &    16.2 &   1  & \\

  NGC 2456 &   118.5444 &    55.4953 &   107.3 &  9.83 &   0.106 & $-$25.33 &  214 &    & 10.9 &         &   1  & \\
  NGC 2492 &   119.8738 &    27.0264 &    97.8 &  9.60 &   0.109 & $-$25.36 &  243 &273 &  8.7 &    12.6 &   3 B & \\
  NGC 2513 &   120.6028 &     9.4136 &    70.8 &  8.74 &   0.063 & $-$25.52 &  274 &    & 13.9 &    24.0 &   4 B & \\
  NGC 2672 &   132.3412 &    19.0750 &    61.5 &  8.35 &   0.058 & $-$25.60 &  268 &    & 16.9 &    14.3 &   3 B & \\
  NGC 2693 &   134.2469 &    51.3474 &    74.4 &  8.60 &   0.054 & $-$25.76 &  349 &    & 13.7 &    15.4 &   1  & \\
  NGC 2783 &   138.4145 &    29.9929 &   101.4 &  9.32 &   0.082 & $-$25.72 &  301 &254 & 11.8 &    38.2 &   3 B & \\
  NGC 2832 &   139.9453 &    33.7498 &   105.2 &  8.70 &   0.047 & $-$26.42 &  332 &    & 18.2 &    21.2 &   4 B & A779 \\
  NGC 2892 &   143.2205 &    67.6174 &   101.1 &  9.35 &   0.233 & $-$25.70 &  304 &    & 12.5 &    23.3 &   1  & \\
  NGC 2918 &   143.9334 &    31.7054 &   102.3 &  9.57 &   0.053 & $-$25.49 &  258 &223 &  8.8 &    18.9 &   1  & \\
  NGC 3158 &   153.4605 &    38.7649 &   103.4 &  8.80 &   0.036 & $-$26.28 &  343 &300 & 14.2 &    16.1 &   6 B & \\

  NGC 3209 &   155.1601 &    25.5050 &    94.6 &  9.34 &   0.060 & $-$25.55 &  303 &    &  9.0 &    29.4 &   3 B & \\
  NGC 3332 &   160.1182 &     9.1825 &    89.1 &  9.37 &   0.087 & $-$25.38 &  221 &220 & 12.5 &    23.7 &   1  & \\

   \hline
  \end{tabular}
\end{center}
\end{table*}

\begin{table*}
\begin{center}
\text{Table 3, continued}

  \begin{tabular}{rccccccccccll}
   \hline

   Galaxy &  R.A. &  Dec. & $D$   &  $K$  & $A_V$ & $M_K$  & $\sigma^{\rm HL}$ & $\sigma^{\rm NSA}$ & $R_e^{\rm 2MASS}$ & $R_e^{\rm NSA}$ & Env. & Note  \\
          & (deg) & (deg) & (Mpc) & (mag) & (mag) &  (mag) & (km/s)           & (km/s)           & (arcsec)        & (arcsec)       &      &  \\
    (1)   &  (2)  &  (3)  &  (4)  &  (5)  &  (6)  &   (7)  &  (8)             &  (9)             &    (10)         &  (11)          & (12) & (13) \\

   \hline
   \hline

  NGC 3343 &   161.5432 &    73.3531 &    93.8 &  9.57 &   0.331 & $-$25.33 &      &     &  10.6 &       &    1 & \\
  NGC 3462 &   163.8378 &     7.6967 &    99.2 &  9.37 &   0.081 & $-$25.62 &  218 &     &  10.1 &  20.1 &    1 & \\
  NGC 3562 &   168.2445 &    72.8793 &   101.0 &  9.38 &   0.111 & $-$25.65 &  264 &     &   8.6 &       &  3 B & \\
  NGC 3615 &   169.5277 &    23.3973 &   101.2 &  9.45 &   0.049 & $-$25.58 &  259 & 271 &   8.2 &  20.2 &  3 B & \\
  NGC 3805 &   175.1736 &    20.3430 &    99.4 &  9.30 &   0.064 & $-$25.69 &  293 & 295 &   8.3 &  16.5 &   42 & \\
  NGC 3816 &   175.4502 &    20.1036 &    99.4 &  9.60 &   0.052 & $-$25.40 &      & 207 &  10.7 &  32.9 &   42 & \\
  NGC 3842 &   176.0090 &    19.9498 &    99.4 &  9.08 &   0.059 & $-$25.91 &  315 & 291 &  14.1 &  24.2 & 42 B & A1367 \\
  NGC 3862 &   176.2708 &    19.6063 &    99.4 &  9.49 &   0.064 & $-$25.50 &  271 & 260 &  11.1 &  40.0 &   42 & \\

  NGC 3937 &   178.1776 &    20.6313 &   101.2 &  9.42 &   0.117 & $-$25.62 &  309 & 289 &  10.7 &  34.7 & 10 B & \\
  NGC 4055 &   181.0059 &    20.2323 &   107.2 &  9.76 &   0.095 & $-$25.40 &      & 270 &   8.1 &  17.5 &  18  & $=$NGC 4061 \\
  NGC 4065 &   181.0257 &    20.2351 &   107.2 &  9.69 &   0.098 & $-$25.47 &  272 & 283 &   8.8 &  31.0 & 18 B & \\
  NGC 4066 &   181.0392 &    20.3479 &   107.2 &  9.81 &   0.086 & $-$25.35 &      & 253 &  10.0 &  37.0 &  18  & \\
  NGC 4059 &   181.0471 &    20.4098 &   107.2 &  9.75 &   0.079 & $-$25.41 &      & 206 &   9.6 &  33.0 &  18  & \\
  NGC 4073 &   181.1128 &     1.8960 &    91.5 &  8.49 &   0.074 & $-$26.33 &  277 & 292 &  21.4 &  23.0 & 10 B & \\
  NGC 4213 &   183.9064 &    23.9819 &   101.6 &  9.61 &   0.102 & $-$25.44 &  259 & 264 &  11.6 &  33.6 &  4 B & \\
  NGC 4472 &   187.4450 &     8.0004 &   16.7* &  5.40 &   0.061 & $-$25.72 &  289 &     &  53.9 &       &205 B &  $=$M49,Virgo \\
  NGC 4486 &   187.7059 &    12.3911 &   16.7* &  5.81 &   0.063 & $-$25.31 &  336 &     &  41.3 &  48.7 & 205  & $=$M87,Virgo \\
  NGC 4555 &   188.9216 &    26.5230 &   103.6 &  9.17 &   0.044 & $-$25.92 &  350 & 319 &  10.1 &  29.8 &   1  & \\

  NGC 4649 &   190.9167 &    11.5526 &   16.5* &  5.74 &   0.072 & $-$25.36 &  340 &     &  39.8 &  44.1 &  205 & $=$M60,Virgo \\
  NGC 4816 &   194.0506 &    27.7455 &  102.0* &  9.71 &   0.024 & $-$25.33 &  244 &     &  14.5 &  50.6 &  49  &  Coma/A1656 \\
  NGC 4839 &   194.3515 &    27.4977 &  102.0* &  9.20 &   0.028 & $-$25.85 &  285 & 269 &  20.8 &  29.2 &  49  &  Coma/A1656\\
  NGC 4874 &   194.8988 &    27.9594 &  102.0* &  8.86 &   0.025 & $-$26.18 &  279 & 266 &  23.8 &  32.0 &  49  &  Coma/A1656 \\
  NGC 4889 &   195.0338 &    27.9770 &  102.0* &  8.41 &   0.026 & $-$26.64 &  401 & 370 &  19.1 &  33.0 & 49 B &  Coma/A1656 \\
  NGC 4914 &   195.1789 &    37.3153 &    74.5 &  8.65 &   0.037 & $-$25.72 &  225 &     &  12.8 &  31.3 &   1  & $=$NGC 4912 \\
  NGC 5129 &   201.0417 &    13.9765 &   107.5 &  9.25 &   0.078 & $-$25.92 &  277 & 262 &  12.3 &  21.8 &   1  & \\
  NGC 5208 &   203.1163 &     7.3166 &   105.0 &  9.51 &   0.097 & $-$25.61 &      & 252 &   6.8 &  18.3 &  3 B & \\
PGC 047776 &   203.4770 &     3.2836 &   103.8 &  9.73 &   0.076 & $-$25.36 &      &     &   7.9 &  13.2 &  9 B & \\
  NGC 5252 &   204.5661 &     4.5426 &   103.8 &  9.77 &   0.095 & $-$25.32 &  196 &     &   9.3 &  19.8 &   9  & \\

  NGC 5322 &   207.3133 &    60.1904 &    34.2 &  7.16 &   0.038 & $-$25.51 &  236 &     &  26.6 &  20.1 &  8 B & \\
  NGC 5353 &   208.3613 &    40.2831 &    41.1 &  7.63 &   0.035 & $-$25.45 &  290 &     &  14.2 &  27.8 & 12 B & \\
  NGC 5490 &   212.4888 &    17.5455 &    78.6 &  8.92 &   0.073 & $-$25.57 &  288 &     &  10.1 &  19.5 &   1  & \\
  NGC 5557 &   214.6071 &    36.4936 &    51.0 &  8.08 &   0.016 & $-$25.46 &  281 &     &  16.2 &  14.7 &  4 B & \\
   IC 1143 &   232.7345 &    82.4558 &    97.3 &  9.51 &   0.172 & $-$25.45 &      &     &   9.2 &       &  3 B & \\
 UGC 10097 &   238.9303 &    47.8673 &    91.5 &  9.38 &   0.049 & $-$25.43 &  302 &     &   7.6 &  17.2 &  7 B & \\
  NGC 6223 &   250.7679 &    61.5789 &    86.7 &  9.11 &   0.100 & $-$25.59 &      &     &  10.6 &       &  4 B & \\
  NGC 6364 &   261.1139 &    29.3902 &   105.3 &  9.74 &   0.106 & $-$25.38 &  205 &     &   7.7 &  11.5 &   1  & \\
  NGC 6375 &   262.3411 &    16.2067 &    95.8 &  9.42 &   0.334 & $-$25.53 &  220 &     &  10.8 &       &   1  & \\
 UGC 10918 &   264.3892 &    11.1217 &   100.2 &  9.31 &   0.498 & $-$25.75 &      &     &  12.8 &       &   1  & \\

  NGC 6442 &   266.7139 &    20.7611 &    98.0 &  9.59 &   0.239 & $-$25.40 &  240 &     &   9.1 &       &   1  & \\
  NGC 6482 &   267.9534 &    23.0719 &    61.4 &  8.37 &   0.277 & $-$25.60 &  322 &     &  10.1 &       &  3 B & \\
  NGC 6575 &   272.7395 &    31.1162 &   106.0 &  9.56 &   0.172 & $-$25.58 &  306 &     &   9.0 &       &   1  & \\
  NGC 7052 &   319.6377 &    26.4469 &    69.3 &  8.58 &   0.337 & $-$25.67 &  284 &     &  14.7 &       &   1  & \\
  NGC 7242 &   333.9146 &    37.2987 &    84.4 &  8.33 &   0.415 & $-$26.34 &      &     &  27.7 &       & 15 B & \\
  NGC 7265 &   335.6145 &    36.2098 &    82.8 &  8.69 &   0.325 & $-$25.93 &  258 &     &  16.0 &       & 21 B & \\
  NGC 7274 &   336.0462 &    36.1259 &    82.8 &  9.24 &   0.295 & $-$25.39 &  298 &     &  12.6 &       &  21  & \\
  NGC 7386 &   342.5089 &    11.6987 &    99.1 &  9.42 &   0.200 & $-$25.58 &      &     &  11.6 &  38.1 &  3 B & \\
  NGC 7426 &   344.0119 &    36.3614 &    80.0 &  8.82 &   0.337 & $-$25.74 &      &     &  11.2 &       &  4 B & \\
  NGC 7436 &   344.4897 &    26.1500 &   106.6 &  9.01 &   0.250 & $-$26.16 &  352 &     &  19.1 &  25.0 &  8 B & \\

  NGC 7550 &   348.8170 &    18.9614 &    72.7 &  8.91 &   0.375 & $-$25.43 &  255 &     &  12.4 &  28.0 &  3 B & \\
  NGC 7556 &   348.9353 &  $-$2.3815 &   103.0 &  9.25 &   0.097 & $-$25.83 &  268 &     &  16.9 &  26.4 &  4 B & \\
  NGC 7618 &   349.9468 &    42.8526 &    76.3 &  9.04 &   0.609 & $-$25.44 &  298 &     &   9.8 &       & 10 B & \\
  NGC 7619 &   350.0605 &     8.2063 &    54.0 &  8.03 &   0.224 & $-$25.65 &  324 &     &  14.8 &  34.6 & 12 B & \\
  NGC 7626 &   350.1772 &     8.2170 &    54.0 &  8.03 &   0.197 & $-$25.65 &  274 &     &  20.1 &  26.7 &  12  & \\
  NGC 7681 &   352.2287 &    17.3096 &    96.8 &  9.22 &   0.149 & $-$25.72 &  359 &     &  11.4 &   5.5 &   1  & \\

  \hline
\end{tabular}
\end{center}

Notes. Column (1): in order of increasing R.A. 
Column (2): right ascension in degrees (J2000.0).  
Column (3): declination in degrees (J2000.0). 
Column (4): distance. Symbol $*$ indicates SBF distances; others are from group-corrected flow velocities,
as described in Sec~\ref{sec:distance}.
Column (5): ``total'' galaxy apparent $K$-band magnitude from 2MASS XSC (parameter k\_m\_ext). 
Column (6): foreground galactic extinction in Landolt $V$-band \citep{schlaflyetal2011} with reddening relation of \citet{fitzpatrick1999}. 
Column (7): extinction-corrected ``total'' absolute $K$-band magnitude derived from distance in column (4), apparent magnitude in column (5),
and foreground extinction in column (6), using Equation~(1).   
Column (8): central stellar velocity dispersion from HyperLeda. 
Column (9): central stellar velocity dispersion from the NASA-Sloan Atlas. 
Column (10): effective radius from 2MASS, defined in Equation~(\ref{eq:reff}). 
Column (11): optical half-light radius from the NASA-Sloan Atlas. 
Column (12): galaxy environment  -- number of group members in the 2MASS HDC group catalog \citep{crooketal2007};
``B'' indicates brightest group galaxy.
Column (13): additional comments, e.g., alternative names, associated clusters.
\\
\\
$^\dagger$ NGC 545 is a close companion of NGC 547.  It is not listed in 2MASS but is designated the BCG of Abell 194 with $M_V=-22.98$ in \citet{laueretal2007}.
%
\end{table*}

\newpage

\section{}


\begin{figure}

\includegraphics[width=7in]{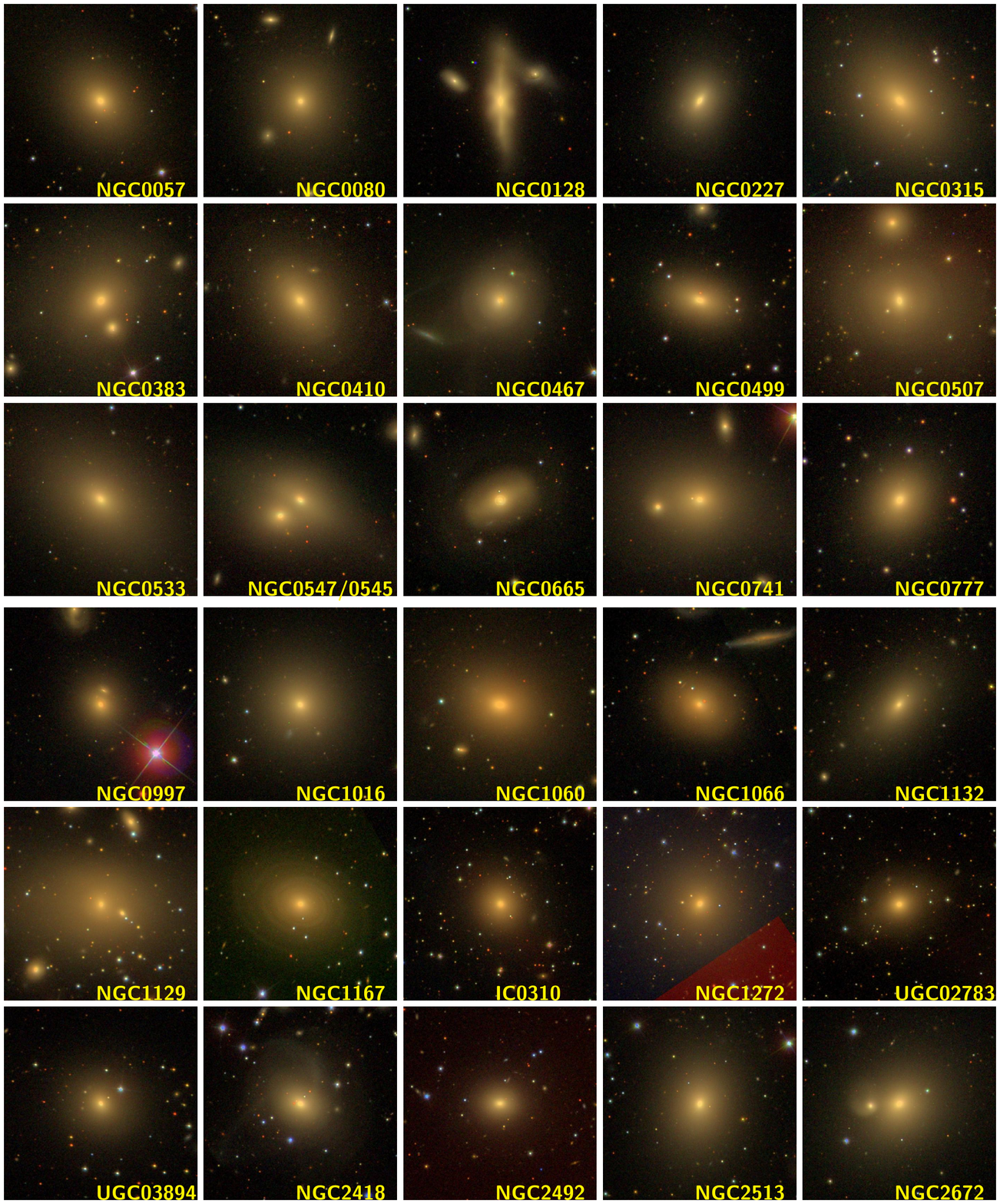}
\end{figure}
\begin{figure}
\includegraphics[width=7in]{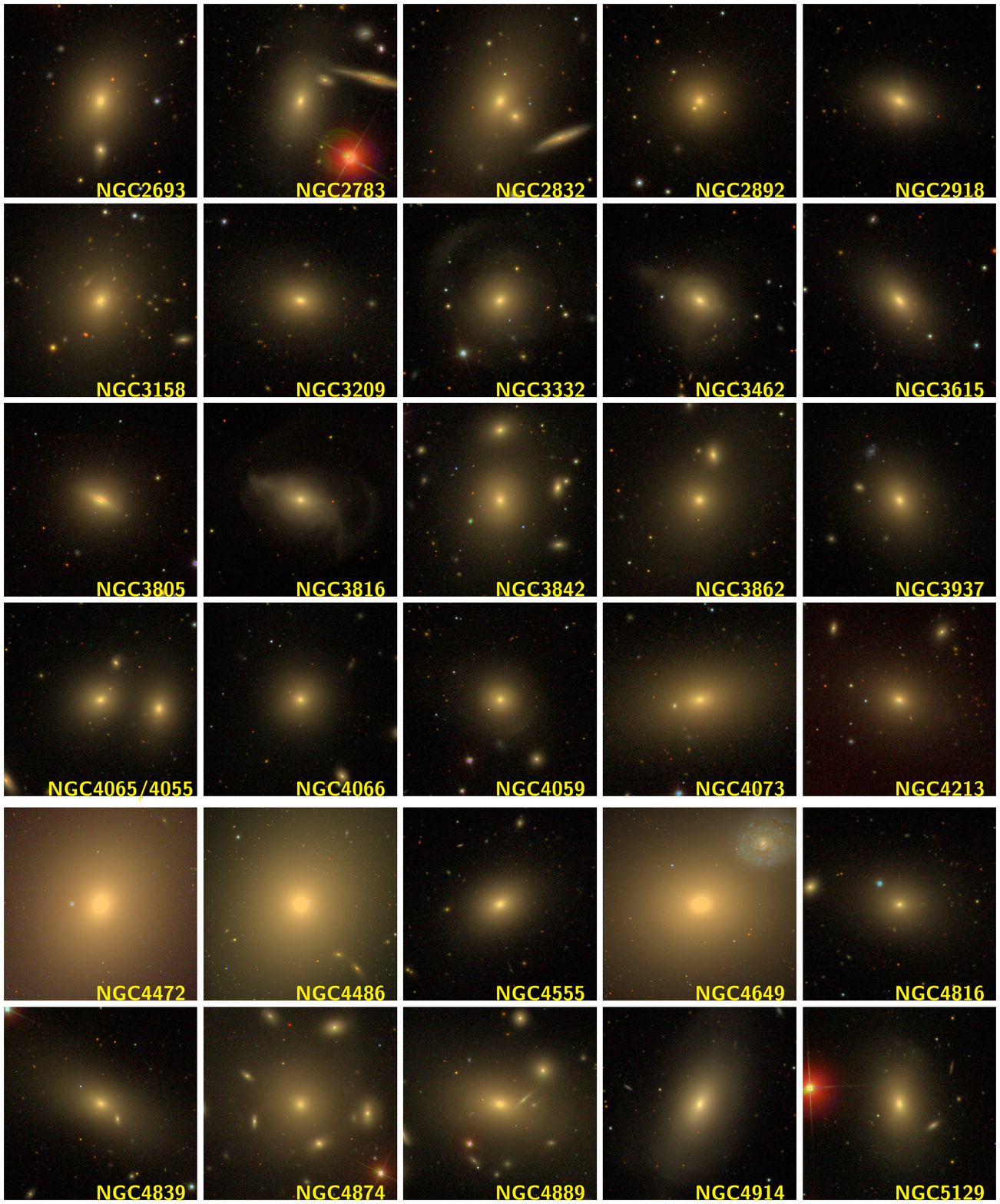}
\end{figure}
\begin{figure}
\includegraphics[width=7in]{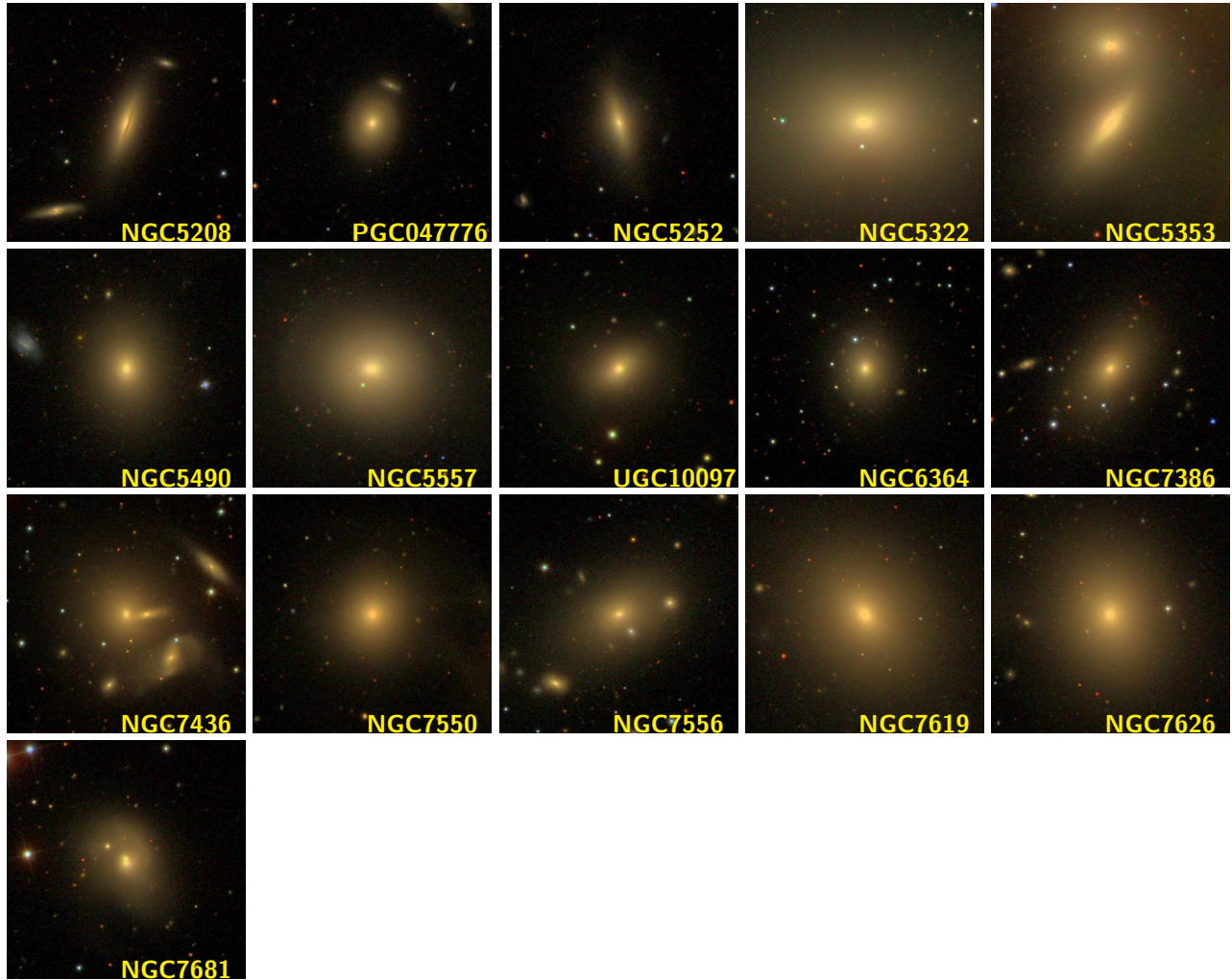}
\vspace{-3in}
\caption{Red-blue-green composite images of SDSS photometry for 78 MASSIVE
  galaxies.  The rest of the sample is not in the SDSS footprint.  Each
  postage stamp shows a 220\arcsec$\times$220\arcsec\ field of view. The
  three exceptions are the Virgo galaxies NGC 4472, NGC 4486, and NGC 4649,
  which are zoomed out to 340\arcsec$\times$340\arcsec. We note that NGC
  4472 was in SDSS but not in NSA.  
}
\label{montage}
\vspace{0.3in}
\end{figure}


\newpage
\bibliographystyle{apj}
\bibliography{massive}
\label{lastpage}

\end{document}